\documentclass[twocolumn,citeautoscript,pra,superscriptaddress,amsmath,amssymb,8pt]{revtex4-1}

\usepackage{hyperref}
\usepackage{graphicx}
\usepackage{gensymb}
\usepackage{color}
\usepackage[dvipsnames]{xcolor}
\usepackage{float}
\usepackage{upgreek}
\usepackage{bm}
\usepackage{amsmath,amssymb}
\usepackage{upgreek}

\newcommand{\ket}[1]{\ensuremath{\left| #1 \right\rangle}}

\graphicspath{{C:/Users/Alex/Desktop/plots/CCPS/}}

\begin{document}

\title{Varied magnetic phases in a van der Waals easy-plane antiferromagnet revealed by nitrogen-vacancy center microscopy}
\author{A. J. Healey}
\thanks{These authors contributed equally to this work.}
\affiliation{School of Physics, University of Melbourne, VIC 3010, Australia}
\affiliation{Centre for Quantum Computation and Communication Technology, School of Physics, University of Melbourne, VIC 3010, Australia}

\author{S. Rahman}
\thanks{These authors contributed equally to this work.}
\affiliation{School of Engineering, College of Engineering and Computer Science, Canberra, ACT 2601, Australian National University}

\author{S. C. Scholten}
\affiliation{School of Physics, University of Melbourne, VIC 3010, Australia}

\author{I. O. Robertson}
\affiliation{School of Physics, University of Melbourne, VIC 3010, Australia}
\affiliation{School of Science, RMIT University, Melbourne, VIC 3001, Australia}

\author{G. J. Abrahams}
\affiliation{School of Physics, University of Melbourne, VIC 3010, Australia}

\author{N. Dontschuk}
\affiliation{School of Physics, University of Melbourne, VIC 3010, Australia}

\author{B. Liu}
\affiliation{School of Engineering, College of Engineering and Computer Science, Canberra, ACT 2601, Australian National University}

\author{L. C. L. Hollenberg}
\affiliation{School of Physics, University of Melbourne, VIC 3010, Australia}
\affiliation{Centre for Quantum Computation and Communication Technology, School of Physics, University of Melbourne, VIC 3010, Australia}	

\author{Y. Lu}
\email{yuerui.lu@anu.edu.au}
\affiliation{School of Engineering, College of Engineering and Computer Science, Canberra, ACT 2601, Australian National University}
\affiliation{Centre for Quantum Computation and Communication Technology, School of Engineering, The Australian National University, Canberra, ACT 2601, Australia}

\author{J.-P. Tetienne}
\email{jean-philippe.tetienne@rmit.edu.au}
\affiliation{School of Physics, University of Melbourne, VIC 3010, Australia}
\affiliation{Centre for Quantum Computation and Communication Technology, School of Physics, University of Melbourne, VIC 3010, Australia}	
\affiliation{School of Science, RMIT University, Melbourne, VIC 3001, Australia}

\begin{abstract}

Interest in van der Waals materials often stems from a desire to miniaturise existing technologies by exploiting their intrinsic layered structure to create near atomically-thin components that do not suffer from surface defects. One appealing property is easily-switchable yet robust magnetic order, a quality only sparsely demonstrated in the case of in-plane anisotropy. In this work, we use widefield nitrogen-vacancy (NV) center magnetic imaging to measure the properties of individual flakes of CuCrP$_2$S$_6$, a multiferroic van der Waals magnet known to exhibit weak easy-plane anisotropy in the bulk. We chart the crossover between in-plane ferromagnetism in thin flakes down to the trilayer, and the bulk behaviour dominated by a low-field spin-flop transition. Further, by exploiting the directional dependence of NV center magnetometry, we are able to observe an instance of a predominantly out-of-plane ferromagetic phase near zero field, in contradiction with expectation and previous experiments on the bulk material. We attribute this to the presence of surface anisotropies arising from the sample preparation process or exposure to the ambient environment, which is expected to have more general implications for a broader class of weakly anisotropic van der Waals magnets.
\end{abstract}

\maketitle 
Recent years have seen intensified interest in intrinsically layered magnetic two-dimensional (2D) van der Waals materials~\cite{Huang2017,Gong2017,Gong2019} as playgrounds for fundamental physics research and as components in next-generation electrical devices, either by themselves or in stacked heterostructures~\cite{Novoselov2016,Jiang2018,Zhang2020a}. For many envisaged applications, a desirable property is to have easily-switchable and robust magnetic order existing down to the monolayer or at least the few-layer limit. Following initial demonstrations of robust out-of-plane, uniaxial ferromagnetism, there has been recent interest in identifying 2D magnets exhibiting easy-plane anisotropy, stemming from a desire to test XY spin models~\cite{Kosterlitz1973} and their potential for enabling easy magnetic switching via the proximity effect in heterostructures~\cite{Liu2013,Cai2019,Purbawati2020}. The existence of XY ferromagnetism in a monolayer was recently demonstrated for the first time  under controlled labratory conditions~\cite{Bedoya-pinto2021}, but questions remain over the stability of such low anisotropy systems in the presence of external perturbations, such as imperfections resulting from the sample preparation process. In this work, we probe the magnetic properties of individual mechanically exfoliated flakes of CuCrP$_2$S$_6$ (CCPS), a van der Waals antiferromagnet known to exhibit weak easy-plane anisotropy in the bulk~\cite{Colombet1982}.

The magnetic structure of CCPS in the bulk form has been established as being characterized by ferromagnetic exchange within a layer, with spins lying in the $ab$ plane, and antiferromagnetic exchange between layers~\cite{Colombet1982}. If it persists down to the few-layer limit, this magnetic order (termed layered or ``A-type" antiferromagnetism) presents opportunities to utilise the ferromagnetic character of extremal layers while producing negligible stray field, making such materials appealing candidates for use in heterostructures relying on an interfacial magnetic exchange interaction~\cite{Jungwirth2016,Ghiasi2021}. This magnetic structure is also appealing in the context of replacing synthetic antiferromagnets with near atomically thin structures~\cite{Duine2018,Seo2021}. Interest in CCPS is further motivated by the copresence of antiferroelectric order~\cite{Maisonneuve1993,Maisonneuve1995,Susner2017,Susner2020}, with the possibility of leveraging a coupling between magnetic and other ferroic orders being mooted~\cite{Lai2019}.

\begin{figure*}[htb!]
\centering
\includegraphics[width=0.95\textwidth]{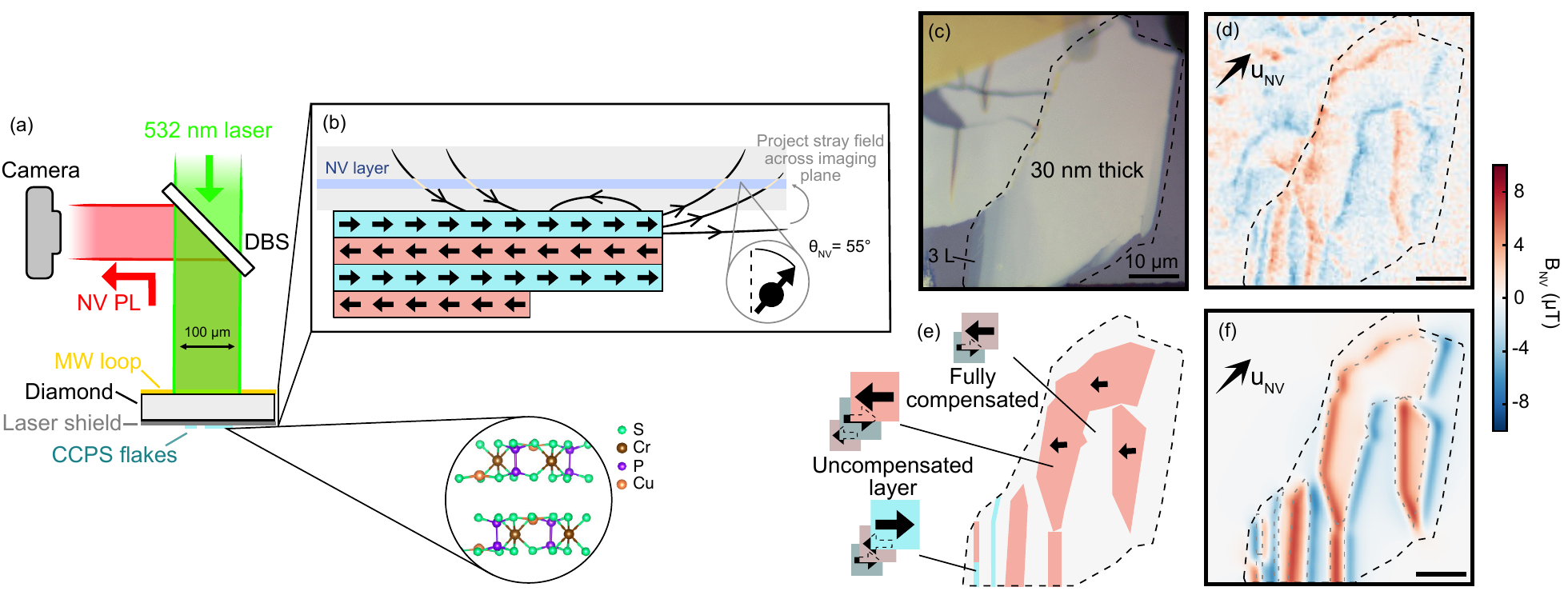}
\caption{\textbf{Imaging in-plane A-type antiferromagnetism with a widefield nitrogen-vacancy center microscope.} (a)~Schematic of measurement setup. The principle of the widefield NV microscope is to image microwave-dependent PL from a well-defined NV sensing layer onto a camera for processing. Circular pull-out: ball-and-stick model of bilayer CCPS in the antiferroelectric phase. DBS: dichroic beam splitter. (b)~Schematic illustration of NV-based stray field sensing of CCPS. CCPS flakes placed onto the diamond surface can be imaged through the stray field that emanates from them, as projected onto a chosen NV axis across the sensing plane, giving a measured field $B_{\rm NV}$. Only odd-layered regions exhibit net magnetization in the A-type antiferromagnetic configuration. (c)~Optical image of a large CCPS flake (largest section about 30~nm thick). (d)~Experimental $B_{\rm NV}$ map of the region from (c) obtained using pulsed ODMR. (e)~Schematic illustration of the domain structure that matches the main features in (d). (f)~Simulated magnetic image of the region from (c), using the domains illustrated in (e). Good agreement with (d) is found when a uniform areal magnetization $M_{\rm s} =3.5~\mu_B$/nm$^2$, evidencing the A-type antiferromagnetic structure.}
\label{fig1}
\end{figure*}

The combination of A-type antiferromagnetism and easy-plane magnetic anisotropy makes CCPS relatively rare among presently identified van der Waals materials, and invites comparison to CrCl$_3$, which is easily-switchable and has been shown to maintain this kind of magnetic order down to the bilayer and monolayer~\cite{Mcguire2017,Bedoya-pinto2021,Ho2019,Klein2019,Wang2019}, and the more strongly anisotropic CrSBr~\cite{Telford2020,Lee2021} which has recently been used to generate spin current in graphene~\cite{Ghiasi2021}. The multiferroicity of CCPS and its weak in-plane anisotropy make it another appealing candidate to join this existing sandbox.

While the antiferroelectric and structural aspects of CCPS are well-documented~\cite{Maisonneuve1993,Maisonneuve1995,Susner2017,Cajipea1996,Levcenko2003,Moriya2005,Susner2020}, the magnetic characterisation of this material is less developed. Previous studies have been limited to bulk measurements of powdered samples~\cite{Colombet1982,Maisonneuve1995,Kleemann2011} or many stacked single crystal flakes~\cite{Kleemann2011,Lai2019}, meaning the properties of individual few-layer flakes have not yet been investigated. Crucially, as the magnetic properties of CCPS have only been established in the bulk, it is uncertain how uncompensated layers, expected to act as effective ferromagnetic monolayers but which may be subject to surface anisotropy and degradation due to air exposure, will behave. More broadly, the question of whether the CCPS monolayer or uncompensated surface layer magnetic properties match the expectation given by the properties of the bulk material is of more general significance for a range of weakly anisotropic easy-plane (anti)ferromagnets desirable for use in van der Waals heterostructures but not yet measured in the ultra-thin limit. 

Here we employ the recently developed technique of widefield nitrogen-vacancy (NV) center magnetometry~\cite{Rondin2014,Levine2019,Scholten2021} to provide quantitative magnetic images of individual few-layer CCPS flakes with sub-micron spatial resolution~\cite{Broadway2020}. In contrast to scanning NV magnetometry~\cite{Thiel2019}, the widefield methodology trades superior spatial resolution for high measurement throughput, thereby allowing extensive magnetic characterisation of a number of different CCPS flakes and under a variety of conditions. Our widefield NV microscope is embedded within a closed-cycle cryostat (base temperature 5\,K) and is equipped with a superconducting vector electromagnet that provides uniform magnetic fields of up to 1~T in any direction~\cite{Lillie2020}. As depicted schematically in Fig. \ref{fig1}(a), a green (532~nm) laser, required for NV initialization and readout, illuminates a spot of diameter 100~$\upmu$m and the resulting NV fluorescence, from which NV spin state information is inferred, is imaged onto a camera to obtain quantitative magnetic images. An NV sensing layer designed to allow rapid imaging near the optical diffraction limit ($\approx 500$~nm) is created via ion implantation of a type-Ib diamond substrate (details in SI Sec. \ref{si:diamond})~\cite{Healey2020}. 

CCPS flakes of various thicknesses were exfoliated from a bulk, commercially-sourced CCPS crystal and transferred directly to the diamond substrate (see details in SI Sec. \ref{SI: CCPS prep}). We use a pulsed optically detected magnetic resonance (ODMR) protocol to measure the stray field $B_{\rm NV}$ emanating from uniformly magnetized domains as projected along a chosen NV axis (oriented at an angle of 55$\degree$ to the surface normal) and across the sensing plane. For an A-type antiferromagnet with easy-plane anisotropy, the stray field is dominantly detected along the edges of domains lying normal to the direction of magnetization~\cite{Fabre2021} as depicted in Fig. \ref{fig1}(b). Ferromagnetic signal should only originate in regions with an uncompensated odd layer and thus the magnetization is expected to be either zero or that corresponding to a monolayer at low fields. 

Figure~\ref{fig1}(c) shows an example optical image of thin CCPS flakes placed onto the diamond surface. The thickness of the flakes was measured using atomic force microscopy (AFM) and phase shift interferometry (PSI) (see SI), which show that this flake is about 30~nm thick (40-50 layers; bulk-like), with some thinner regions with thickness down to three layers. Figure \ref{fig1}(d) shows a magnetic map ($B_{\rm NV}$) of this region obtained using our widefield NV microscope. In the thick region, extended magnetic domains are visible as well as areas exhibiting zero net magnetization, consistent with odd-layer and even-layered stacks respectively.  Magnetic signal is also visible in the thinner region, although the dimensions of these stripe-like domains are on the order of our spatial resolution and so some definition is lost and the magnitude of the stray field detected is artificially reduced. Nevertheless, all stray field detected is of the same order of magnitude agreeing with the expectation of A-type magnetic order. Indeed, the $B_{\rm NV}$ map can be readily reproduced in simulation [Fig.~\ref{fig1}(f)] by defining a series of uniformly magnetized domains representing odd-layered regions, sketched in Fig.~\ref{fig1}(e). Most domains are seen to be oriented towards the left of the page, matching the direction set by a ``training field" $B_{\rm tr}$=0.1\,T applied in-plane at the cryostat base temperature. The bias field used to facilitate the ODMR experiment (an ``applied field" $B_{\rm ap}=6$~mT along the NV axis) is weak enough to not affect the magnetization, however some domains are oriented oppositely due to an earlier magnetization process.

\begin{figure}
\centering
\includegraphics[width=0.47\textwidth]{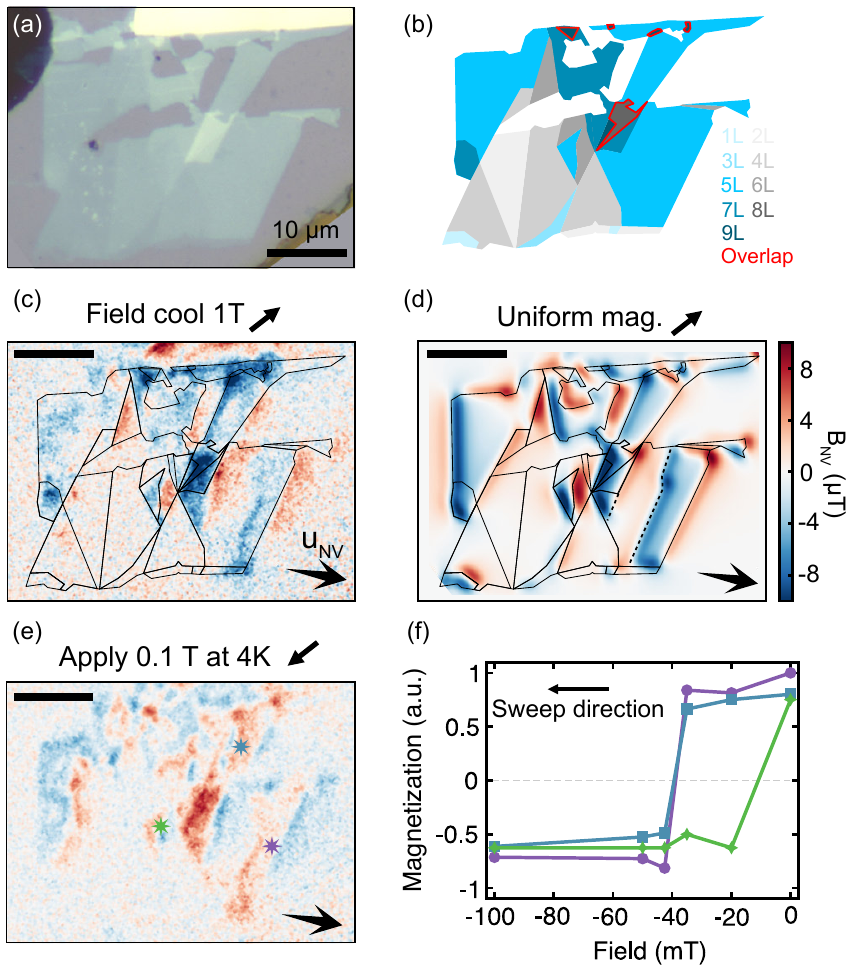}
\caption{\textbf{In-plane ferromagnetism of uncompensated monolayers.} (a)~Optical image of thin CCPS flakes. (b)~False color image of the region from (a) depicting the thickness of the flakes as measured by PSI (see SI). Shades of blue are odd-layer flakes expected to exhibit net magnetization, shades of grey are even-layered regions whose magnetic sublattices should fully compensate one another, and red outlines are areas with overlapping flakes where the antiferromagnetic coupling may not be sustained throughout the stack. (c)~Magnetic field map recorded at a field of $B_{\rm ap} = 6$~mT applied along the NV axis following a field cooling procedure with $B_{\rm}=1$~T applied in the direction indicated. The black outline shows the optically visible flake borders, with magnetic signal seen to originate in the odd-layered stacks. (d)~Simulated $B_{\rm NV}$ signal taking $M_{\rm s}=3.5~\mu_{\rm B}$/nm$^2$ in the direction indicated. Agreement is obtained with (c) with the exception of monolayer and trilayer regions. Dotted lines indicate domain boundaries included to match experimental image, lying along cracks visible in AFM (see SI). (e)~As (c) but following the application of a switching field of 0.1~T in the opposite direction at the base temperature of 4~K. Almost all domains are seen to switch magnetization. (f)~Evolution of magnetization of the regions marked in (e) under switching fields of different magnitude. Coercive fields of 40~mT are typical with a spread depending on flake-dependent anisotropy and size effects.}
\label{fig2}
\end{figure}

Although the reconstruction of magnetization of in-plane magnets from their stray field is prone to error~\cite{Broadway2020a}, in principle we can measure the spontaneous magnetization  $M_{\rm s}$ of the CCPS monolayer by comparing our measurements with simulation~\cite{Fabre2021}. Taking into account known parameters such as the NV-CCPS standoff and the measurement's spatial resolution (see details in SI \ref{si: sims}), a value of $M_{\rm s} = 3.5$~$\mu_B$/nm$^2$ provides a good match. This value is short of the theoretical value obtained from a first-principles calculation, $M_{\rm s} = 9.6$~$\mu_B$/nm$^2$, however these values are broadly consistent with one another given the sources of error. These include artificially reduced signal due to delocalized NV PL contributions (up to a factor of two in Ref.~\cite{Fu2020}), the finite measurement temperature~\cite{Maisonneuve1995}, and possible sample degradation during sample transfer. In light of this, our measured value represents a lower bound of the actual $M_{\rm s}$ and we use the theoretical $M_{\rm s}$ in the analysis to come, only using the magnitude of stray fields measured comparatively. 

Figure~\ref{fig2}(a) shows an optical image of a region containing few-layer flakes of various thicknesses. Through PSI measurement (see SI), the thickness of these flakes is measured as summarised in the false color image Fig.~\ref{fig2}(b). Magnetizing the flakes by field cooling under a field $B_{\rm tr} = 1$~T and measuring again under a bias field $B_{\rm ap} = 6$~mT, it can be seen in Fig.~\ref{fig2}(c) that stray field features arise within odd-layered regions and at their borders, as expected. The magnitude of the measured field does not vary with flake thickness, and so we again identify the magnetic signal with that of uncompensated CCPS monolayers in odd-layered stacks. The exception to this is for regions that are identified as being comprised of two overlapping flakes [highlighted in red in Fig. \ref{fig2}(b)], where a standoff or lattice mismatch may exist, in both cases precluding significant antiferromagnetic exchange and resulting in a signal corresponding to two uncompensated monolayers. Figure~\ref{fig2}(d) shows the simulated magnetic signal assuming all odd-layered flakes exhibit an identical magnetization in the direction of the training field, with good agreement with experiment obtained when accounting for flake defects visible in AFM [see SI, domain boundaries defined by dotted lines in Fig.~\ref{fig2}(d)]. No magnetic signal is observed in the two monolayer regions, however as above we can identify signal from the trilayer region under some conditions (see SI). It is reasonable to suspect that long-range magnetic order may cease to exist when not stabilised by the interlayer exchange (or exist with a dramatically lowered $T_c$), however it also cannot be ruled out that this sample's brief exposure to air could result in sample degradation which could be especially significant for thinner flakes. 

\begin{figure*}[htb]
\centering
\includegraphics[width=\textwidth]{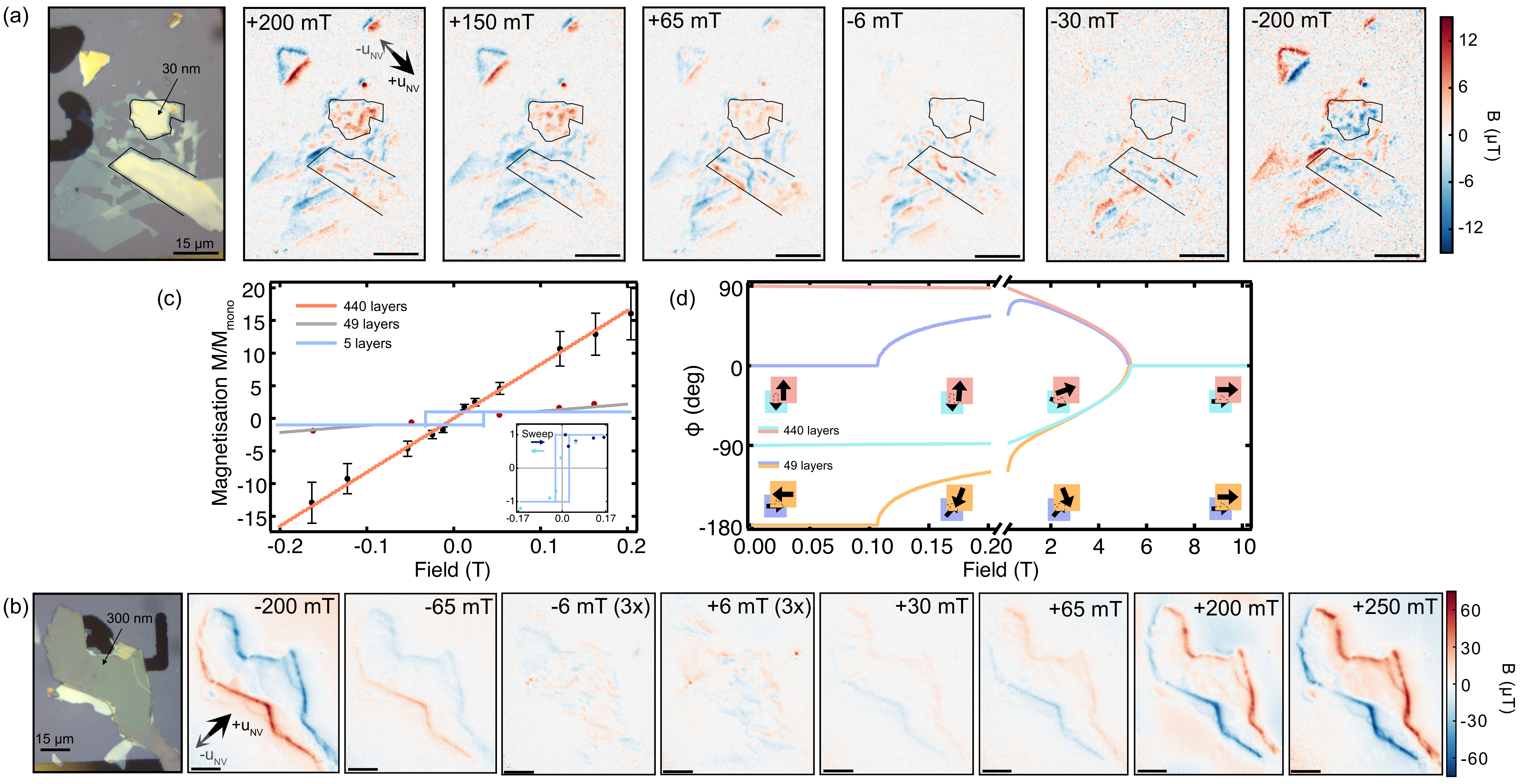} 
\caption{\textbf{Thickness-dependent response to applied fields.} (a)~Series of measurements taken at a range of magnetic fields $B_{\rm ap}$ applied along the NV axis. The optical image (left) shows the region over which these images were taken, which includes the region of Fig. 2. Different behaviours are observed for thin, intermediate, and thick flakes which is evidence of a layer-dependent spin-flop transition. (b)~As (a) but for a thick ($\approx 300$~nm) flake, pictured on the left. The measured signal is seen to vary smoothly with the applied field, consistent with a near-zero-field spin-flop and coherent rotation of sublattice spins towards $B_{\rm ap}$. In all cases a positive $B_{\rm NV}$ field is defined as being a positive projection along the respective $+\hat{u}_{\rm NV}$ axis. (c)~Measured magnetization for 300~nm, 40~nm, and 5-layer flakes (black points, red points, and inset respectively) versus the in-plane component of $B_{\rm ap}$ (note that the labels in (a) and (b) are the total magnitude of $B_{\rm ap}$). Solid curves are theoretical predictions made using a micromagnetic calculation for the respective data sets (see text). At minimum qualitative agreement is found in each case. (d)~Plot of the angle $\phi$ the sublattice spins are predicted to make with the in-plane projection of $B_{\rm ap}$ (chosen to lie along $\phi = 0$), as given by the micromagnetic calculation of (c). Over the range of fields accessible in our experiment ($<0.2$~T), only a small net canting towards the field is predicted, however the moments will saturate to a ``forced ferromagnetic" configuration at fields $>5$~T.}
\label{fig: MH}
\end{figure*}

The ferromagnetic character at the cryostat base temperature of 5~K of the odd-layered flakes was investigated by attempting to switch the magnetization with the application of increasingly large fields up to $B_{\rm ap} = 0.1$~T in the direction opposite to the initial field cooling, with a magnetic image of the region being recorded in between. The extremal image is shown in Fig.~\ref{fig2}(e), showing almost all regions magnetized in the opposite direction to Fig.~\ref{fig2}(c), and Fig.~\ref{fig2}(f) shows a pseudo-hysteresis curve constructed for representative flakes, as inferred by linecuts (extended data in SI). It can be seen that the magnetization of most flakes switches under the application of a coercive field of 35-40~mT, although there is a spread due to variable flake-specific anisotropy. These results confirm that odd-layered CCPS behaves as an easily-switchable ferromagnet, potentially making the proposed magnetoelectric switching viable~\cite{Lai2019}. 

A more direct method of measuring the evolution of the magnetization is to apply the field directly along a chosen NV axis and image at that field. Imaging at larger fields allows the bulk magnetization dynamics of CCPS to be revealed as sufficiently strong magnetic fields will cause the fully-compensated state to be a higher energy configuration than ones resulting in net magnetization. In the following, we assume that only the field applied in-plane contributes. In Fig.~\ref{fig: MH}(a) we begin with the region from Fig.~\ref{fig2}, expanded to include bulk-like flakes as well as few-layer flakes (optical image left), and show a series of images taken over a range of applied fields. We see that, consistent with the coercive field measurements, the structure of the magnetic signal in thin flakes is seen to align along the applied field past 30~mT. Features in thin flakes do not change dramatically for higher fields than this, although the edges of two 40-50 layer regions (marked in black, henceforth referred to as ``intermediate thickness") become more well-defined and increase in magnitude slightly compared to the low-field case (a factor of two between 6 and 200~mT). As this change appears to occur suddenly between 65 and 150~mT, it is consistent with a finite-field spin-flop transition taking place \cite{Stryjewski1977}.

More strikingly, above 65~mT clear magnetic signals emerge from thick ($>50$ layers) flakes shown in Fig \ref{fig: MH}(a) that are not visible (or barely visible) at low fields. We interpret this as thick stacks undergoing a spin-flop transition near zero field and thus exhibiting a net magnetic moment in the direction of the applied field, which varies smoothly with this field's magnitude. 

We can reproduce the behaviour observed in both thick and intermediate thickness flakes using a simple micromagnetic model that solves for the lowest-energy configuration of two magnetic sublattices with magnetizations $\bm{m}_1$ and $\bm{m}_2$, previously applied successfully on CrCl$_3$~\cite{Wang2019}. Here we assume that the intralayer ferromagnetic exchange interaction is strong enough for all spins within a layer to be locked with one another, so that it suffices to consider a linear, antiferromagnetically coupled chain with an exchange coupling $J_{\rm AFM}$ between adjacent spins~\cite{Wang2019}. We also assume that magnetic anisotropy within the plane is negligible, so that in full generality the total energy of a system of $N$ spins is given by:

\begin{equation}
E_{\rm tot} = -\sum_i^N  M_{\rm s} B \cos{(\phi_i)} + \sum_j^{N-1} J_{\rm AFM} \cos{(\phi_j-\phi_{j+1})}
\label{micromag}
\end{equation}
where the $i^{\rm th}$ spin in the chain makes an angle $\phi_i$ with the external magnetic field of magnitude $B$. Note that this equation can be simplified to considering only two angles $\phi_1$ and $\phi_2$ since in the ground state all spins in the chain belong to one of two magnetic sublattices. For small $B$, the second term dominates and the A-antiferromagnetic order is maintained, however at larger fields a spin-flop transition is favourable. This simple model predicts that the spin-flop threshold should be $N$-dependent; being near zero for even-layered flakes as the Zeeman terms from the two sublattices cancel out in the antiferromagnetic configuration, and scaling inversely with $N$ in the odd-layered case. This behaviour has been studied extensively in CrCl$_3$~\cite{Wang2019}, and qualitatively matches the above observations. 

Performing the same measurements on a large, thick flake [Fig.~\ref{fig: MH}(b)], measured to be 300~nm thick, or up to 440 layers, tells a similar story: the complex stray field structure present at low field quickly disappears as the applied field exceeds 30~mT. The increase in magnetization over the range of accessible fields (up to 250~mT) appears steady, again indicating a smooth rotation of spins from a low-field spin-flop up to saturation at a much higher field.

By taking linecuts across the measurements in Fig.~\ref{fig: MH}(b) the total magnetization can be inferred, relative to the signal previously ascribed to the CCPS monolayer. This data is summarised in Fig.~\ref{fig: MH}(c), along with data from one of the intermediate thickness flakes in Fig.~\ref{fig: MH}(a). The solid curves are theoretical predictions from the micromagnetic model Eqn. \ref{micromag}, taking the theoretical $M_{\rm s}$, values for $N$ inferred from AFM measurements, taking the average coercive field $\approx 40$~mT measured earlier, and leaving $J_{\rm AFM}$ as a fitting parameter. For both sets of data a value $J_{\rm AFM} = (1.2 \pm 0.6)\times 10^{-4}$~J/m$^2$ produces good agreement (error analysis in SI), which is slightly larger than the value coarsely predicted by mean field theory $J_{\rm AFM}/k_B= 1$~K per f.u. ($J_{\rm AFM}=4.3\times 10^{-5}$~J/m$^2$)~\cite{Colombet1982}. This discrepancy is not large, especially considering that variations in stacking patterns due to the exfoliation process have been shown to increase antiferromagnetic exchange by an order of magnitude compared to the bulk in similar systems~\cite{Klein2019}. Figure~\ref{fig: MH}(d) shows the angles the sublattice spins are predicted to make with respect to an in-plane field applied at $\phi = 0$ corresponding to the theoretical curves in Fig.~\ref{fig: MH}(c), as well as a schematic representation of the rotation. Over the range of fields probed in this work only a small rotation from the perpendicular orientation ($<3 \degree$) is predicted for the 440-layer flake, however this is sufficient to produce a large magnetic signal compared to the monolayer. Saturation to a forced ferromagnetic state is predicted to occur at around 5~T in both cases. 

\begin{figure*}
\includegraphics[width=0.9\textwidth]{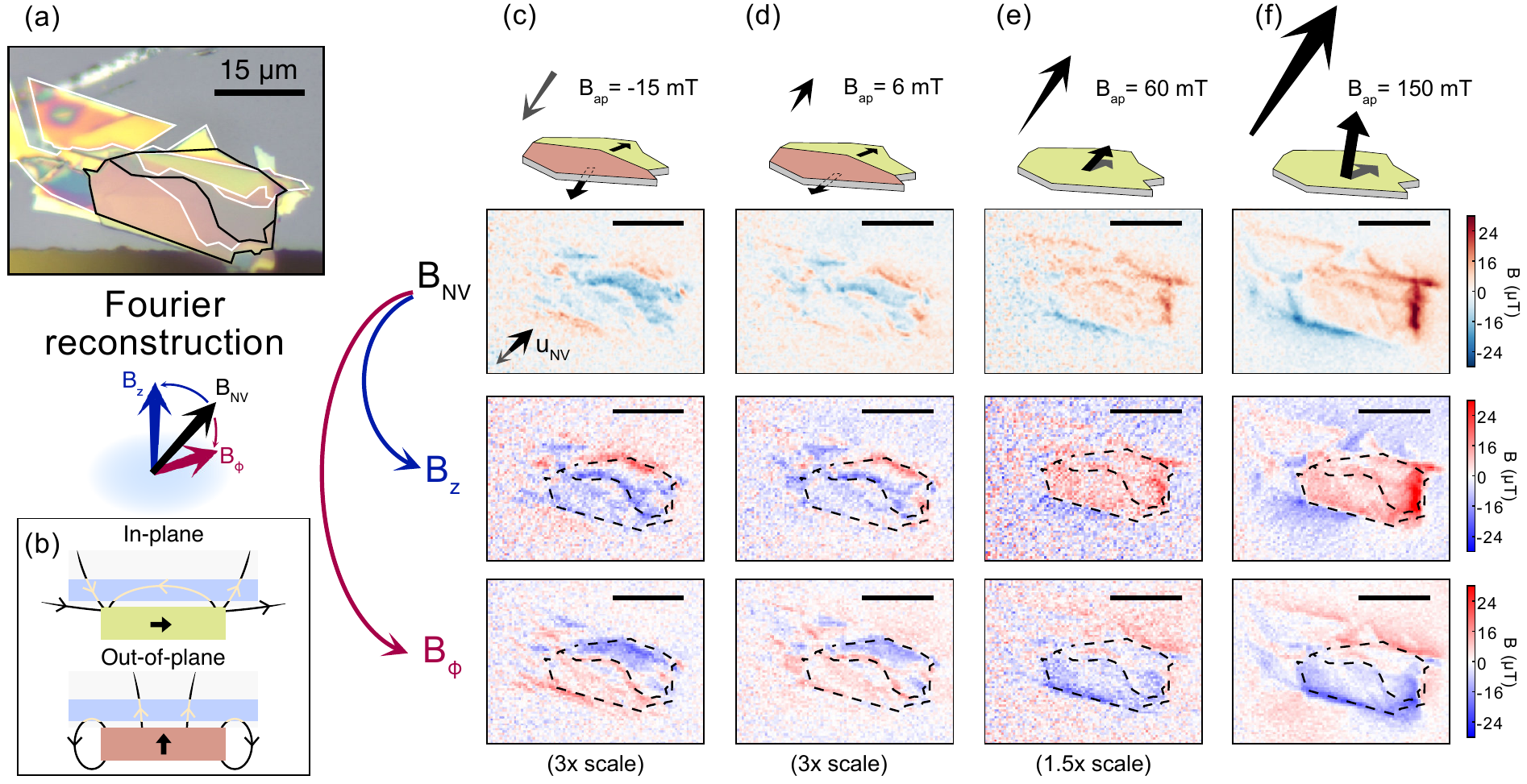}
\caption{\textbf{Observation of an out-of-plane ferromagnetic phase.}~(a) Optical image of 200~nm thick CCPS flake. White outline shows main visible boundaries, some of which are visible in high-field $B_{\rm NV}$ images; black outline shows those that correspond to the main low-field domains. (b)~Schematic showing the direction of stray field from uniformly magnetized domains across an NV layer (blue) for the in- and out-of-plane cases. Within domains (away from edges), the direction of the stray field (roughly speaking) reflects the direction of magnetization.~(c) From top to bottom: schematic representation of inferred magnetic domain structure of the largest part of the flake at a field $B_{\rm ap} = -15$~mT; experimental $B_{\rm NV}$ image of the region shown in (a); magnetic field projection along the $z$ axis ($B_z$) obtained via Fourier reconstruction from the raw $B_{\rm NV}$ image; reconstructed field projection along the in-plane $\phi=45\degree$ direction ($B_{\phi}$). (d), (e), (f)~As (c) but under fields of $B_{\rm ap}=6$~mT, $B_{\rm ap}=60$~mT, and $B_{\rm ap}=150$~mT respectively.}
\label{fig5}
\end{figure*}

Until now, we have assumed that the magnetizations are wholly confined to the $ab$ plane, despite sometimes significant out-of-plane magnetic field projections. In reality, however, bulk CCPS is known to be a weakly anisotropic system~\cite{Colombet1982,Qi2018} and so it is expected that out-of-plane magnetization could be observed under strong applied fields and that surface anisotropies may become relevant in stabilising out-of-plane phases at low fields. To differentiate between in-plane and out-of-plane magnetization in a thick (200~nm) CCPS flake [optical image shown in Fig.~\ref{fig5}(a)], we use $B_{\rm NV}$ measurements (projected along the well-defined axis $\hat{u}_{\rm NV}$, with components both in- and out-of-plane) to reconstruct the full vector magnetic field using upward propagation~\cite{Casola2018}. To provide a qualitative analysis, we note that, as sketched in Fig.~\ref{fig5}(b), the orientation of the stray field measured at the sensing layer is reflective of the direction of the flake's magnetization. A uniformly magnetized domain will give a field oriented along (or against) the magnetization axis within its extent, while the nonzero projection in the perpendicular directions will dominantly be along its edges.  

The layered antiferromagnetic structure of CCPS conveniently allows us to separate the bulk magnetic properties from an uncompensated surface layer; as demonstrated above, the bulk spin-flop phase is evident when measuring thick flakes at fields $>100$~mT while uncompensated odd layers (if present) will dominate the measured signal at low fields. Therefore, images taken of thick flakes with large domains at low field over a range of fields can provide information about both the bulk and surface behaviours.

Figure~\ref{fig5}(c)-(f) shows a series of raw magnetic images obtained at various applied fields, along with the reconstructed field components $B_z$ and $B_{\phi = 45 \degree}$. Starting with a field of $B_{\rm ap} = -15$~mT [Fig.~\ref{fig5}(c)] we see two distinct magnetic domains in the largest part of the flake. From the Fourier reconstruction, it can be seen that one domain is consistent with in-plane magnetization as expected, while the other produces significant internal $z$ field consistent with an out-of-plane (or at least canted) phase. The inferred net magnetization directions within each domain are summarised in the schematic (top row) and are seen to be retained upon moving to an inverted bias field $B_{\rm ap} = +6$~mT [Fig.~\ref{fig5}(d)], albeit with slightly reduced magnitude. This retention confirms that the out-of-plane magnetization is ferromagnetic in nature, contrary to the expectation of only in-plane ferromagnetic phases being supported by odd-layered CCPS. The existence of this anomalous out-of-plane ferromagnetic phase is problematic for any application reliant on robust in-plane magnetism and highlights the fragility of easy-plane anisotropy in van der Waals materials. The likely origin is a surface anisotropy, possibly arising as a result of outer layer degradation due to brief exposure to air or an interfacial effect from the variable stacking pattern visible in the optical image \cite{Johnson1996,Cheng2022}. Regardless of the mechanism, the existence of such a phase is unlikely to be unique to CCPS: this result may have much broader implications for the suitability of a wider range of weakly anisotropic van der Waals magnets. 

Upon increasing the applied field we see that [Fig.~\ref{fig5}(e),(f)], again, the flake begins to exhibit bulk-like behaviour with a consistent magnetization across both regions smoothly increasing with field. Although there is a significant in-plane component, the Fourier reconstruction additionally reveals that a $z$ component also exists, this time reversibly following the applied field. Treating the data from Fig.~\ref{fig: MH} similarly reveals that the magnetization also eventually cants towards the applied field axis in other thick flakes, (see SI for data and discussion). As discussed in the supplementary information, this represents only a minor correction to the two-dimensional model from earlier and does not alter the main conclusions. Field reconstructions presented in the SI also confirm that the field maps from Figs. \ref{fig1} and \ref{fig2} are consistent with in-plane magnetization as claimed.

To conclude, we have probed the magnetic properties of individual CCPS flakes over a range of thicknesses and magnetic fields. We confirm that the in-plane A-type antiferromagnetic order is maintained down to at least the trilayer, and that extended ferromagnetic domains can exist throughout the stack even for flakes of over 40 layers. We have shown that the intrinsic magnetocrystalline anisotropy in CCPS is small, leading to a low-field spin-flop transition in the bulk and easy magnetic switching of flakes with uncompensated layers (coercive field $\approx 40$~mT at 5~K). The robust, low-coercivity ferromagnetic order found to exist in few-layer CCPS leaves open the possibility of switching via electrostatic gating if magnetoelectric coupling can be established~\cite{Lai2019}. Additionally, however, it was found that the weak in-plane anisotropy also allows small perpendicular anisotropies to become significant in stabilising out-of-plane spin configurations. Through Fourier reconstruction of individual field components from our magnetic images, we were able to observe an instance of an out-of-plane ferromagnetic phase in CCPS not hereto reported in the literature. 

We speculate that this configuration could have arisen as the result of a surface anisotropy, whose magnitude may vary from flake to flake due to the level of damage to the surface layer, allowing the uncompensated monolayer to exhibit perpendicular anisotropy while the bulk properties remain relatively unchanged. This interpretation implies that sustaining in-plane magnetic order in monolayer CCPS (or indeed similar low-anisotropy monolayers) may be difficult, although our measurements on small monolayer CCPS flakes were inconclusive.  

These results highlight the challenges remaining in identifying van der Waals materials suitable for use in next-generation electronic and spintronic devices, where exfoliation down to the monolayer or bilayer is frequently desirable but may result in altered properties compared to those of the bulk material. Widefield NV microscopy, as showcased in this work, looms as a critical characterization platform in this context as it offers the potential for quantitative, directional imaging with high spatial resolution and is operable over a wide range of conditions. If future work allows the relationship between material properties and magnetic anisotropy to be understood, the magnetism in such systems can be controlled, allowing more reliable operation and potentially the use of a wider range of magnetic states. 
\section*{Acknowledgements}
This work was supported by the Australian Research Council (ARC) through grants CE170100012, DP190101506, and FT200100073. A.J.H. is supported by an Australian Government Research Training Program Scholarship. S.C.S gratefully acknowledges the support of an Ernst and Grace Matthaei scholarship. 

\bibliographystyle{apsrev} 
\bibliography{CCPS_final}

\pagebreak
\onecolumngrid
\begin{center}
\textbf{\large Supplementary Information}
\end{center}
\setcounter{equation}{0}
\setcounter{section}{0}
\setcounter{figure}{0}
\setcounter{table}{0}
\setcounter{page}{1}
\makeatletter
\renewcommand{\theequation}{S\arabic{equation}}
\renewcommand{\thesection}{S-\Roman{section}}
\renewcommand{\thefigure}{S\arabic{figure}}
\section{Diamond sample details}
\label{si:diamond}
The diamond samples used for this work were (100)-oriented, type-Ib high-pressure high-temperature (HPHT) synthesised substrates purchased from Delaware Diamond Knives. To create a dense NV sensing layer, the diamonds were irradiated with 100~keV $^{12}$C$^-$ ions with a fluence of 10$^{12}$ ions/cm$^2$. The substrates have a native nitrogen density of approximately 100~ppm, and the ion implantation has the effect of incorporating lattice vacancies to increase the N to NV conversion ratio from near-zero to between 1 and 5\% within the implanted region when followed by a high-temperature anneal (a ramped sequence culminating at 1100~$\degree$C~\cite{Tetienne2018}). This defines a sensing layer approximately 100~nm thick 100-200~nm from the diamond surface. Further details and characterization of this process is given in Ref.~\cite{Healey2020}.  

On top of the diamond, an Al grid (80~nm) and an Al$_2$O$_3$ layer (80~nm) were deposited by photolithography and atomic layer deposition respectively. This coating facilitates flake location (correlating the grid as seen in NV PL with photographs of as-deposited flakes), acts as a laser shield to limit sample heating during illumination, and protects the diamond surface from plasma treatments designed to aid flake adhesion (see Sec. \ref{SI: CCPS prep}).

This sensor design was chosen to allow spatial resolution near the optical diffraction limit while maximizing the NV PL collected and thus the magnetic sensitivity~\cite{Healey2020}. The spatial resolution is approximately given by 2$d_{\rm SO}$ where $d_{\rm SO}$ is the maximum standoff between the NV sensing layer and the target flakes~\cite{Abrahams2021}. As the combined standoff given by the mean NV depth from the diamond surface and the Al grid is 250-300~nm, the minimum achievable spatial resolution is about 500~nm. 
\section{CCPS preparation and characterization}
\label{SI: CCPS prep}
\subsection{CCPS preparation}
Thin flakes of CCPS were fabricated by the well-proven mechanical exfoliation method~\cite{Huang2015}. Scotch tape was used to exfoliate commercially purchased bulk crystals in different thicknesses. First, the tape containing CCPS was put into a viscoelastic PDMS stamp and gently peeled off. An ultrahigh resolution optical microscope (Nikon) was utilised to identify thin layers to bulk samples on PDMS based on the color contrast. The diamond substrate was then prepared for transferring 2D flakes. It was thoroughly cleaned in an ultrasonic bath of acetone, isopropanol and deionized water in sequence to remove substrate residues. Subsequently it was subjected to an oxygen plasma treatment to remove possible ambient adsorbates from the surface~\cite{Huang2015}. The Al/Al$_2$O$_3$ coating protected the diamond surface and NV layer during this process. A three-axis transfer stage equipped with a micro-actuator and a microscope was used to facilitate the transfer of CCPS flakes from PDMS onto designated sections of the diamond substrates. During the transfer process, the sample stage was mildly heated to 60-65$\degree$C to help remove any trapped air molecules between the sample and substrate interface. This method is effective in obtaining intact and ultraclean surfaces in addition to achieving contamination free samples~\cite{Rahman2021}. Using this method, we were able to obtain CCPS samples from bulk down to the monolayer with large surface area. Raman spectroscopy was used to confirm the presence of CCPS. The thickness of the samples was measured with atomic force microscopy (AFM) and phase shifting interferometry (PSI). At all times following fabrication, samples were stored and maintained in a N$_2$ cabinet with inert atmosphere, with only intermittent exposure to air. In addition, sample preparation was carried out in shortest possible time to help mitigate the risk of sample degradation.
\subsection{CCPS identification}
\begin{figure}
\includegraphics[width=0.5\textwidth]{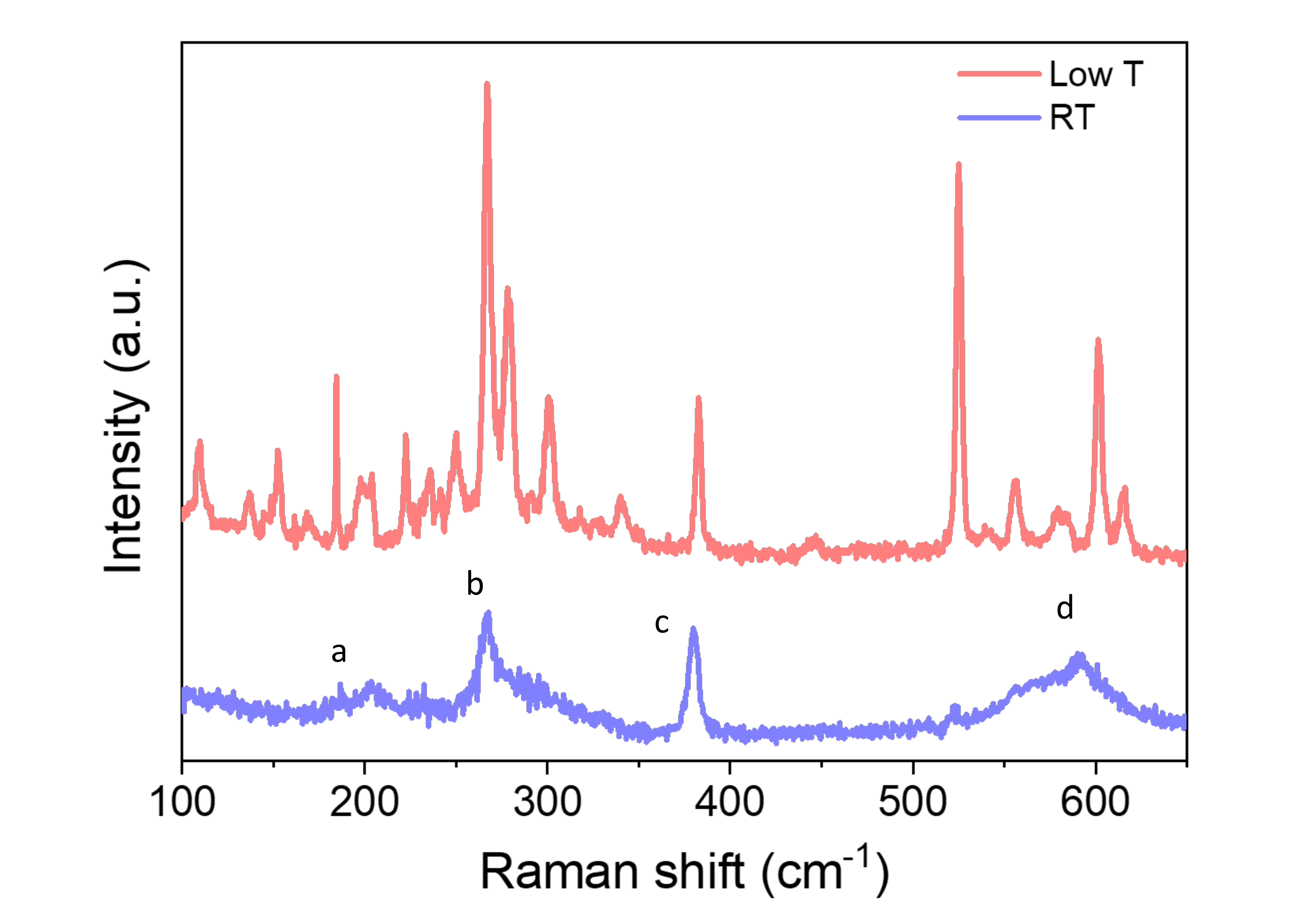}
\caption{CCPS characterization using Raman spectroscopy. Measured Raman spectrum of CCPS at room temperature (RT) and 77~K (low T) showing the Raman-active modes and change in phase transition.}
\label{sifig:raman}
\end{figure}
Raman spectroscopy is a powerful and sensitive optical method to identify 2D materials for their unique non-linear interaction. Vibrational modes of any 2D crystals provide distinct Raman signatures based on their crystal symmetry and lattice structure~\cite{Cong2020}. Figure~\ref{sifig:raman} shows the Raman spectrum of CCPS taken at room temperature and 77\,K. Several modes are observed at low temperature while only four modes can be characterized at room temperature. This response is typical of a pure CCPS crystals as shown in earlier work~\cite{Susner2020}. Four distinct Raman modes, labelled a-d, can be identified based on the direction of lattice vibration. Only the first peak corresponds to the E$_g$ type signalled by in-plane vibration whereas the rest is associated with out of plane A$_g$ type vibration. The [P$_2$S$_6$]$^{4-}$ block within the anionic sublattice of CCPS can undergo various lattice derangement upon thermal or laser excitation. Major effects include rotation and translation of the PS$_3$ group; in addition symmetric stretching of P-P bond and deformation also causes strong perturbation on the vibrational modes of PS$_3$ group. This results in distinct Raman modes at 204~cm$^{-1}$, 266~cm$^{-1}$, 378.5~cm$^{-1}$ and 595~cm$^{-1}$ labelled a-d (Fig. \ref{sifig:raman})~\cite{Poizat1985,Sourisseau1983}. CCPS also exhibits a clear phase transition to the antiferroelectric state below 145~K which is also visible in Raman spectroscopy~\cite{Susner2020}. The low-temperature spectrum demonstrates numerous additional peaks which can be correlated with infrared sensitive phonon modes. This becomes prominent when the crystal undergoes phase transitions and there is significant reduction in the crystal inversion symmetry. Hence the Raman spectra recorded confirms the presence of CCPS crystal.

A Horiba LabRAM system equipped with a confocal microscope, a charge-coupled device Si detector, and a 532~nm diode-pumped solid state laser as the excitation source was used for the Raman measurements. An objective lens was used to focus the laser light on the surface of the sample. The illuminated spot on the samples is estimated to have a diameter of 2~$\upmu$m. The spectral response of the entire system was determined with a calibrated halogen-tungsten light source. For low temperature measurements, the sample was placed on a microscope compatible Linkam chamber connected to a temperature controller; liquid nitrogen was used as the coolant.  
\subsection{CCPS surface topography and thickness}
\begin{figure}
\includegraphics[width=0.5\textwidth]{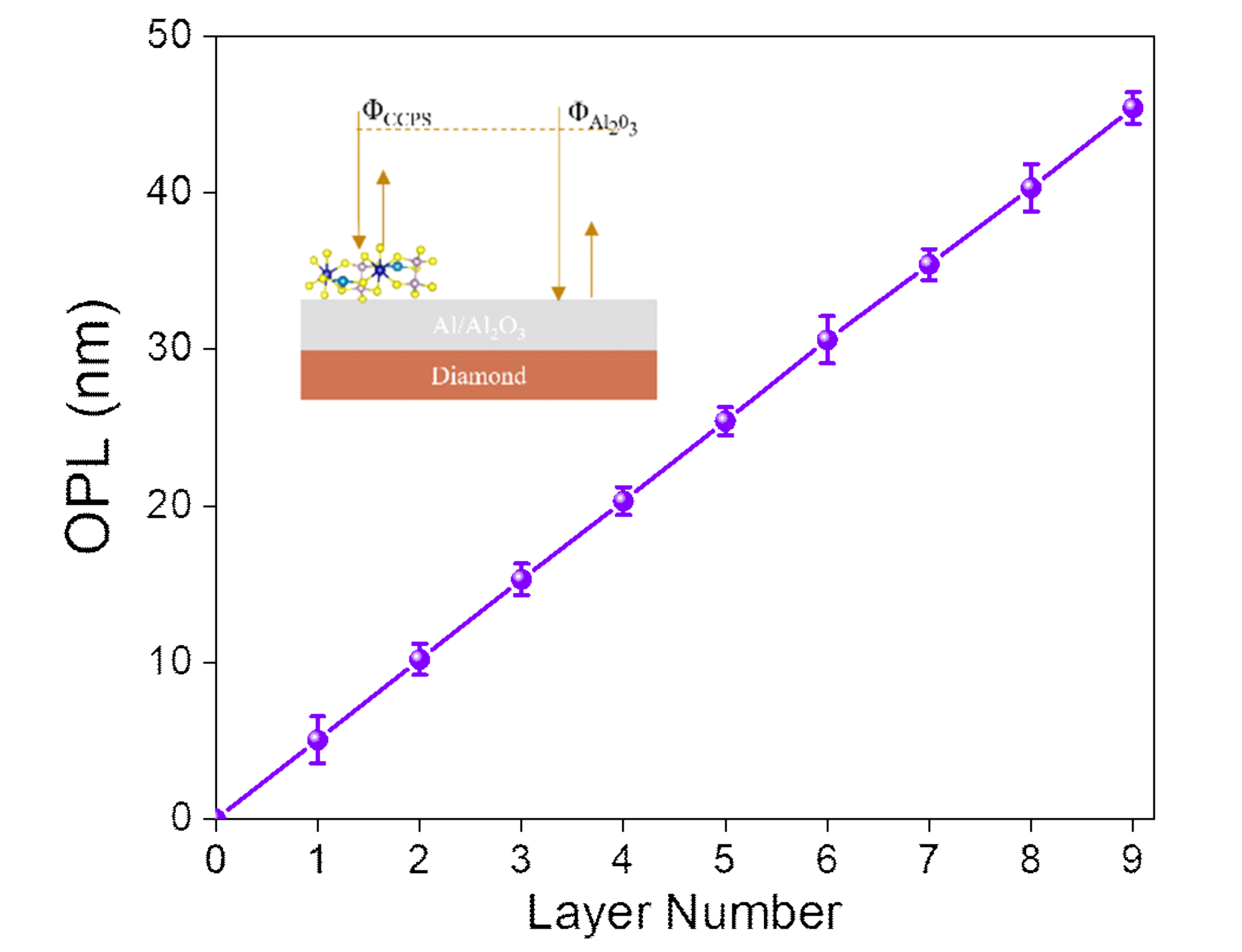}
\caption{Statistical data of the OPL values from PSI for thin CCPS samples. Large number of CCPS samples were characterized for the statistical measurements. Inset is the schematic plot showing the PSI-measured phase shifts of the reflected light from the CCPS flake and the Al$_2$O$_3$ substrate ($\Phi_{\rm Al_2O_3}$).}
\label{sifig:OPL}
\end{figure}
\begin{figure}
\includegraphics[width=0.8\textwidth]{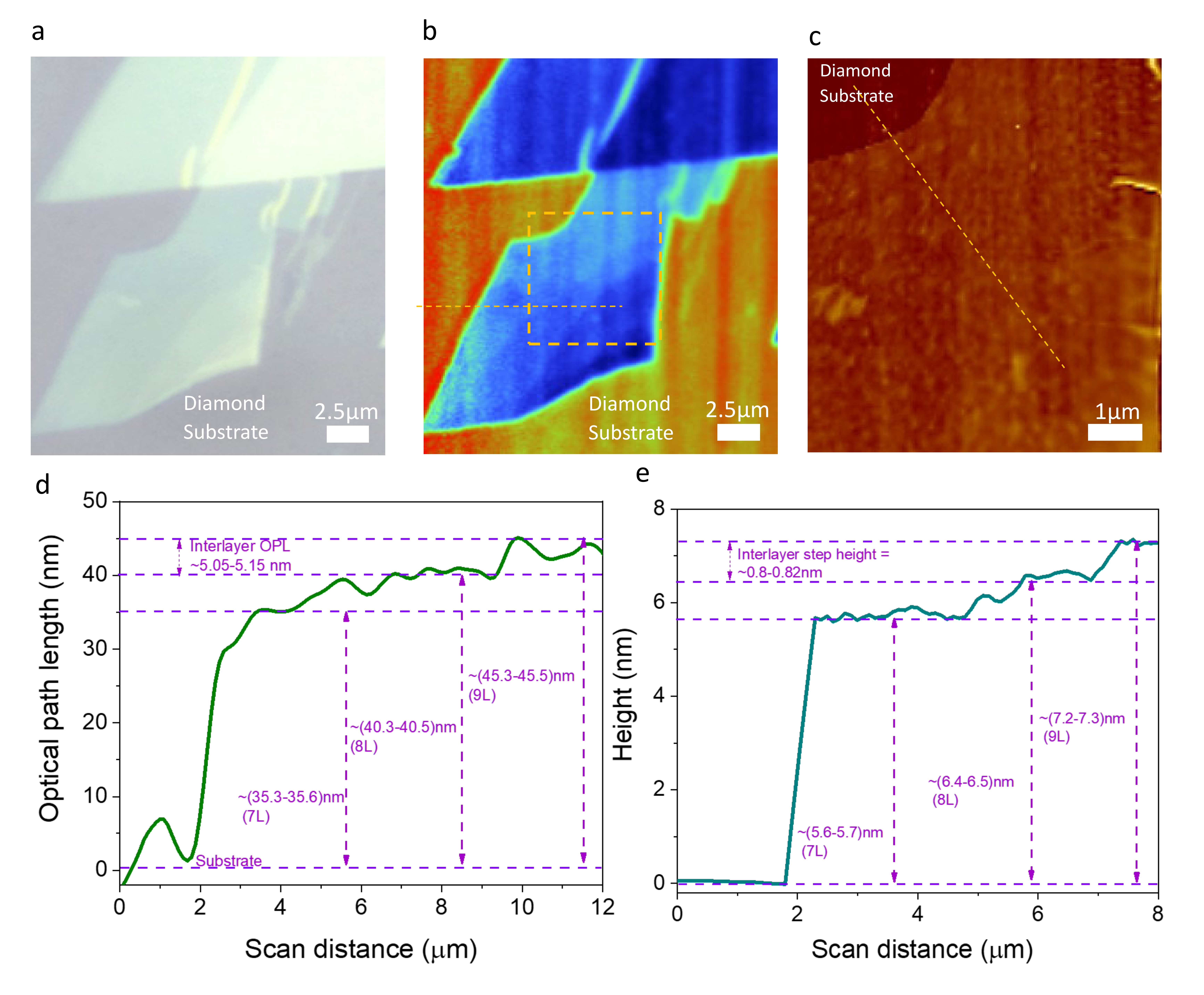}
\caption{(a) Optical microscopy image of a typical CCPS sample with seven, eight and nine layers on diamond substrate. (b) Corresponding PSI image and (c), AFM image of the sample showing the layer contrast. (d) PSI measured optical path length (OPL) profile along the orange dashed line shown in (b); OPL of 7L, 8L and 9L are measured to be 35.3~nm, 40.4~nm and 45.3~nm respectively; interlayer OPL is calibrated to be 5.05-5.15~nm. Corresponding AFM height profile of the section marked in blue dotted rectangle. (e) AFM height profile along the orange dashed line marked in (c). Step height clearly signifies the interlayer difference is about 0.8~nm which is an excellent match with previous report on similar materials~\cite{Liu2016}.}
\label{sifig:OPLAFMlinecuts}
\end{figure}
\begin{figure}
\includegraphics[width=0.7\textwidth]{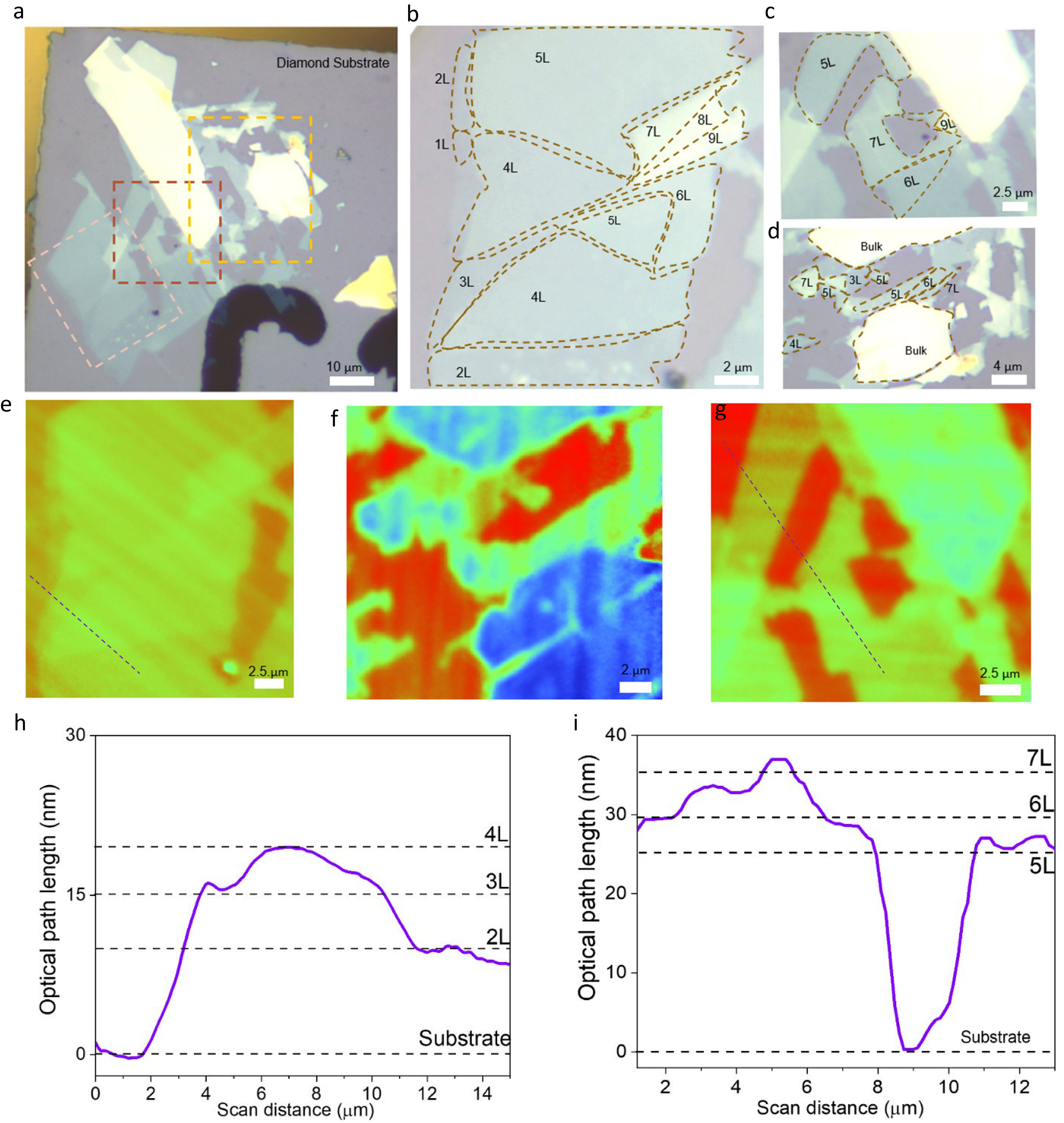}
\caption{(a) Optical image of the thin region from Fig. 2 indicating different thickness present in the sample. (b)-(d). Zoomed in image of gold, red and orange boxes respectively. Layer number of each region are outlined and marked. (e)-(g). High resolution PSI images showing the variation in contrast of different layers. (h),(i) Line cuts taken across purple dotted lines in (e) and (g) respectively, showing the layer differences.}
\label{sifig:psiC1}
\end{figure}
\begin{figure}
\includegraphics[width=0.6\textwidth]{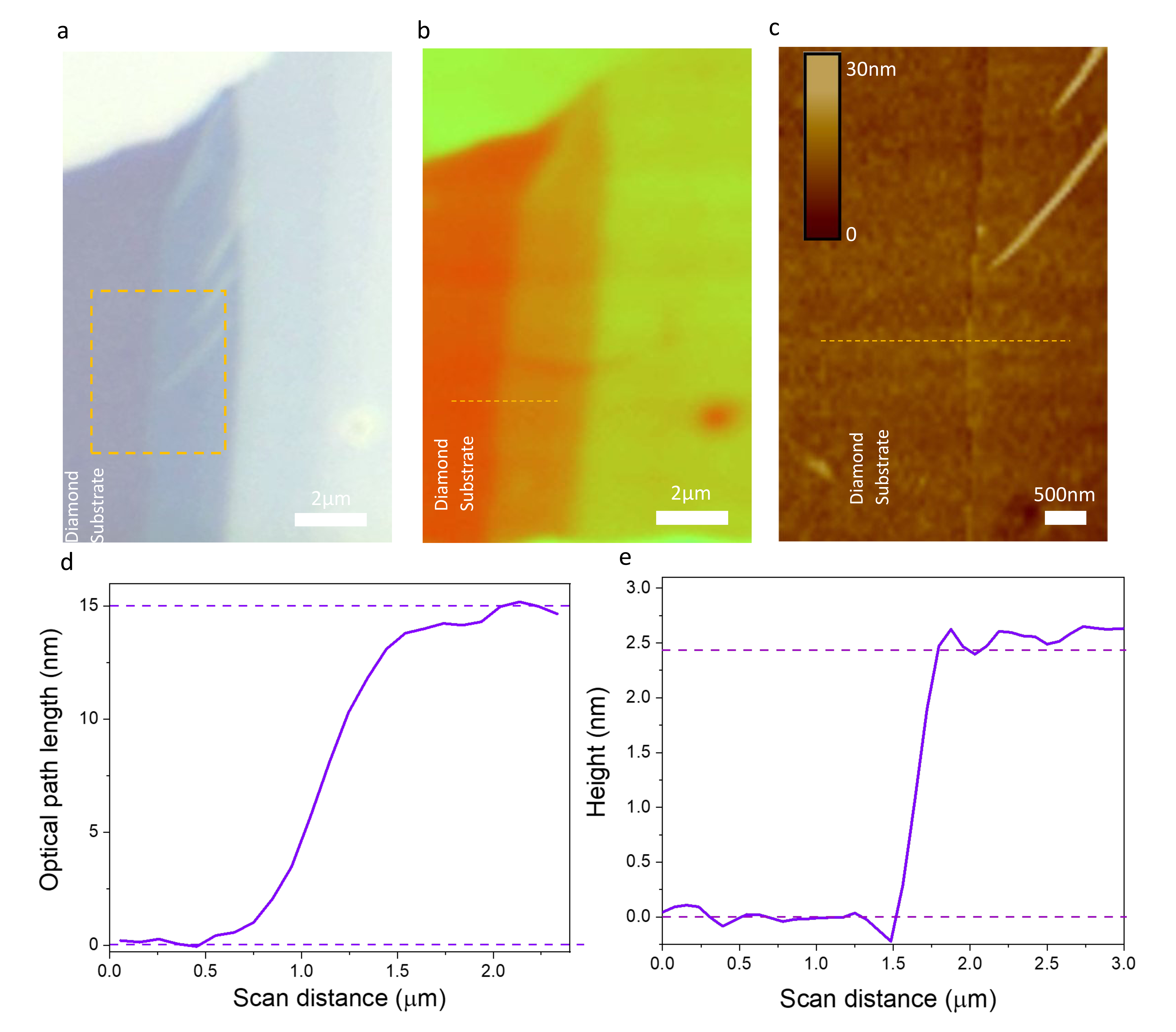}
\caption{(a) Optical microscopy image of thin section of Fig. 1 flake. (b) PSI image of the region from (a). (c) AFM image of the orange box in (a). (d),(e) OPL and AFM line profile taken across the dotted orange line in (b) and (c) respectively. Both PSI and AFM are in good agreement with each other, signifying a three layer sample.}
\label{sifig:trilayer}
\end{figure}
\begin{figure}
\includegraphics[width=0.6\textwidth]{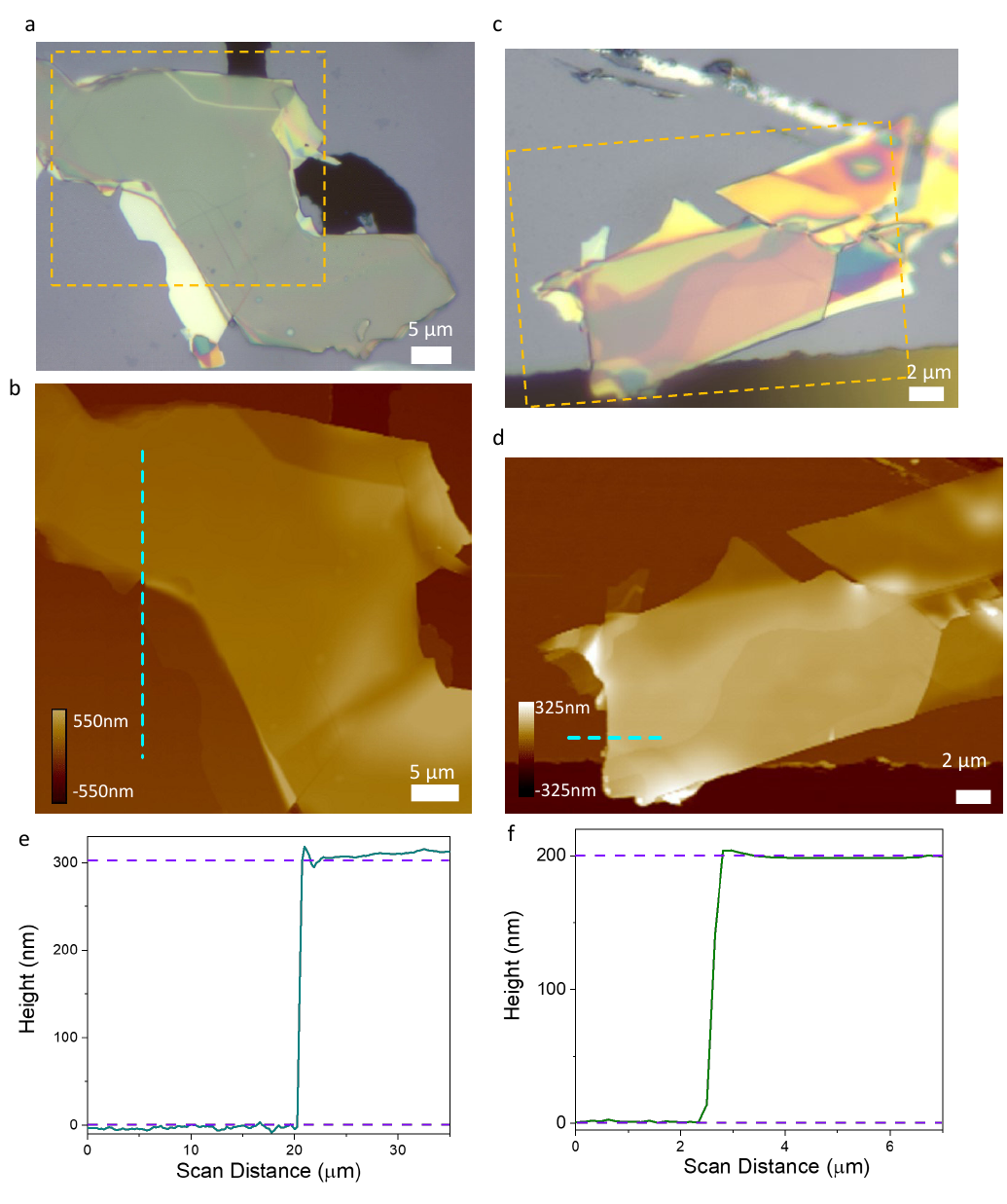}
\caption{Optical Microscope and AFM images of (a),(b) thick flake shown in Fig. 3 (main text) and (c),(d) flake shown in Fig. 4 (main text) respectively. (e),(f) Cross sectional line profile taken across the blue dotted line in (b) and (d) respectively. The flakes are about 300~nm and 200~nm thick respectively.}
\label{sifig:thickafm}
\end{figure}

The thickness of CCPS flakes investigated for this work was measured using both phase shifting interferometry (PSI) and atomic force microscopy (AFM). Phase shifting interferometry has been used to measure the optical path length (OPL) of 2D layers. OPL can be deduced from the relation ${\rm OPL}_i= \frac{\lambda}{2\pi}\left(\phi_i-\phi_{\rm sub}\right)$, where $\lambda$ represents the wavelength of the light used for excitation (i.e. 535~nm), $\phi_i$ and $\phi_{\rm sub}$ signifies the measured phase shift of the reflected light signal from the sample and the substrate (Fig.~\ref{sifig:OPL} inset) respectively~\cite{Yang2016}. It has been demonstrated that PSI is a highly sensitive and powerful tool to characterize atomically thin materials, since it has a high resolution of 0.1~nm in the detection of the surface profile. Figure~\ref{sifig:OPLAFMlinecuts} shows the layer-dependent OPL of CCPS measured by PSI, compared with an AFM measurement of the same region. OPL between each layer was found to be 5.05-5.15~nm. 7L, 8L and 9L CCPS have 35.3~nm, 40.4~nm and 45.3~nm OPL while their AFM heights (5.6~nm, 6.4~nm and 7.2~nm respectively) shows an excellent match with previous reports on similar class of compounds~\cite{Liu2016}. The relationship between OPL and layer number (Fig.~\ref{sifig:OPL}) was confirmed using AFM. Layer number could thus be quickly determined over a large area by the measured OPL values. More examples of surface topography of the thin flakes examined in the main text are shown in Fig.~\ref{sifig:psiC1} and \ref{sifig:trilayer}. However for the thicker samples (beyond 10L) examined in this work, AFM was used to measure sample thickness. Figure~\ref{sifig:thickafm} shows AFM images for the thick flakes studied in Fig. 3 and 4 of the main text. In the case of the Fig. 3 region [Fig.~\ref{sifig:thickafm}(a),(b),(e)], this measurement was used to inform the micromagnetic model Fig. 3(c),(d). For this purpose, it was important to measure the flake thickness over the flattest region where the height measured by AFM is most representative of the total amount of CCPS (noting that stacking faults are possible throughout the stack and there is visible buckling elsewhere on the flake); this was also the region used to infer the flake magnetization through linecuts of the corresponding $B_{\rm NV}$ images for the same reason.  

All optical path length measurements were obtained using a phase shifting interferometer (Veeco NT9100). Bruker Icon Model was used for AFM measurements. Peak force tapping or scanasyst air mode is used to directly measure the surface topography and the same tip was used to complete all scanning related to this work. Geometry of AFM tip: triangular; tip radius: 2~nm; frequency: 70~KHz; length: 115~$\upmu$m; width: 25 µm, spring constant: 0.4~Nm$^{-1}$; drive amplitude: 122.38~mV with amplitude set point 250~mV.

\section{NV measurement setup}
All NV measurements were carried out on a cryogenic widefield microscope, described in detail in Ref.~\cite{Lillie2020}. The closed-cycle cryostat (Attocube attoDRY1000) has a base temperature of less than 4~K, but the sample temperature is estimated to be closer to 5~K (as quoted in the main text) under typical ODMR measurement conditions. Magnetic fields of up to 1~T in any direction can be applied using a superconducting vector magnet (Cryomagnetics). A 532~nm laser (Laser Quantum Ventus) allows NV initialisation and readout and pulse control is provided by a fibre-coupling acousto-optic modulator (AAOpto MQ180-G9-Fio). The laser is passed through a beam scrambler (Optotune LSR-3010) to remove optical interference patterns that reduce the illumination uniformity, before being focussed using a low-temperature microscope objective (Attocube LT-APO/VISIR/0.82) to obtain roughly even laser illumination over a region covering most of the $\sim$100~$\upmu$m field of view (see Fig.~\ref{SI_odmr}(b) below). The laser power density at the NV layer, following losses along the optical path, is estimated to be around 1~kW/cm$^2$. NV PL is collected through the same objective, filtered (709/167~nm) and imaged onto a sCMOS camera (Andor Zyla). 

The diamond sample is glued onto a glass coverslip, patterned with a gold microwave resonator to facilitate NV spin state driving, which is then connected to a printed circuit board. A signal generator (Rohde \& Schwarz SMB100A) provides the microwave signal, gated by a switch (Mini-Circuits ZASWA-2-50DR+) and amplified (Mini-Circuits HPA-50W-63). Pulse sequences (including synchronising with camera acquisition) were programmed onto and controlled by a SpinCore PulseBlaster ESR-PRO 500 MHz card.
\section{NV measurement details}
All magnetic measurements presented were obtained using pulsed optically detected magnetic resonance (ODMR). A 20~$\upmu$s laser pulse (chosen to achieve a balance between NV initialization and readout contrast) was followed by a microwave $\pi$ pulse ($\approx 300$~ns) of a given frequency. This sequence was then repeated to fill the 30~ms camera exposure. A second camera exposure follows with a no-microwave sequence to act as a reference, against which the first exposure can be normalised. This was then repeated over a desired range of frequencies to build up an ODMR spectrum. Following sufficient signal acquisition (typically several hours to obtain a low-noise image; thousands to tens of thousands total sweeps), a magnetic image can be obtained by extracting the ODMR spectrum at each camera pixel and fitting the resonance frequencies. 

\begin{figure}
\includegraphics[width = 0.75\textwidth]{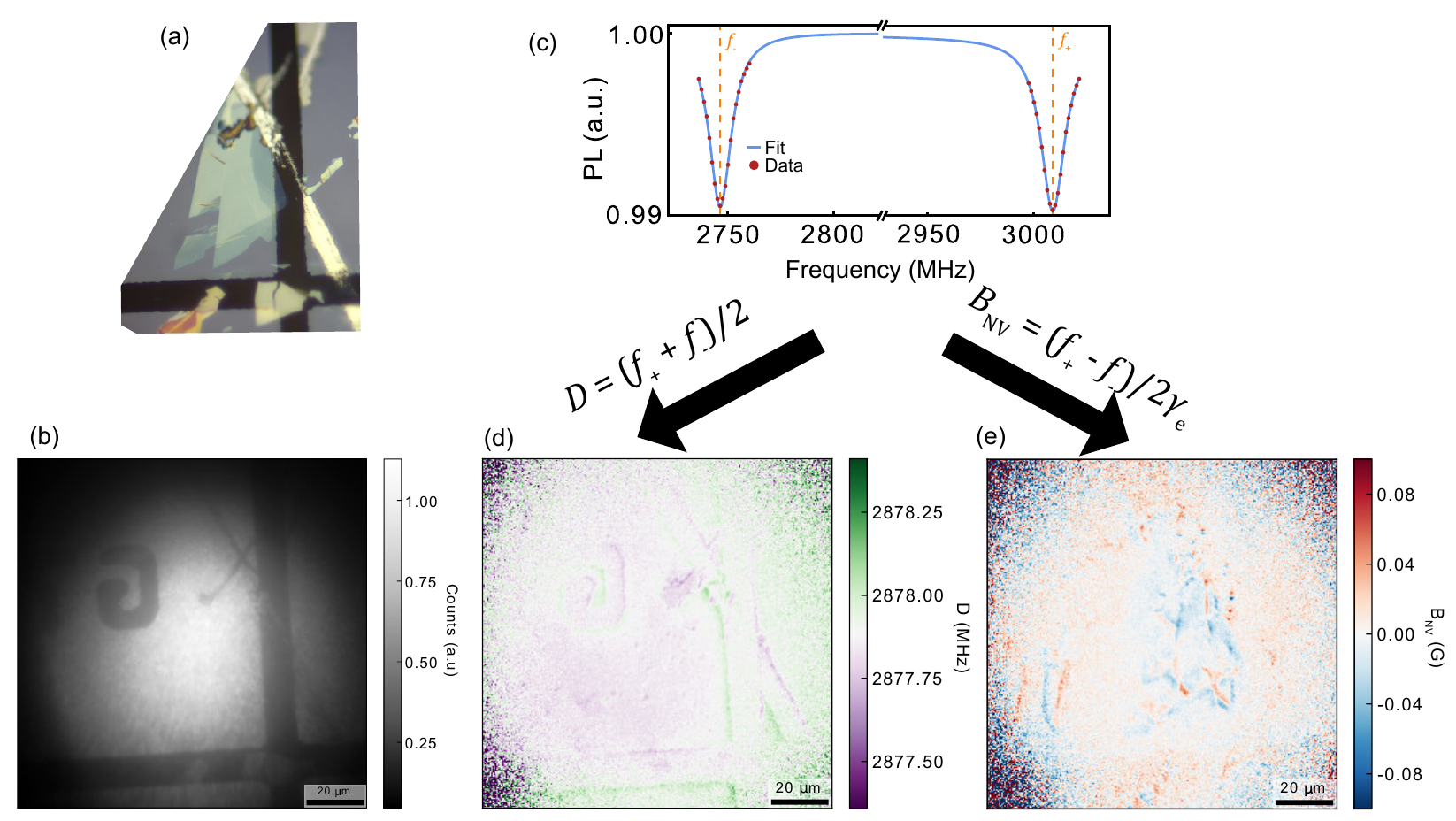}
\caption{(a) Optical image of thin CCPS flakes on top of the diamond sensor. The Al/Al$_2$O$_3$ grid is visible as well as a scratch. (b) NV PL image of the full measurement FOV. Higher PL is measured in regions where Al has been deposited due to reflection. (c) Example NV ODMR spectrum taken under an applied field of 60~G. The blue line is a Lorentzian fit to the data. (d) Map of the zero field splitting parameter $D$, which shows variations in crystal strain across the FOV due to the Al deposition and scratching of the diamond surface. (e) $B_{\rm NV}$ map showing magnetic features from the CCPS flakes without influence from the strain inhomogeneity.}
\label{SI_odmr}
\end{figure}

A typical measurement taken under a low bias field ($|B|\leq 30$~mT) was obtained by applying that bias along a chosen NV axis (for a (100) diamond these axes point at 45$\degree$, 135$\degree$, 225$\degree$, and 315$\degree$ from the $x$ axis, and 54.7$\degree$ from the $z$ axis). For these low bias fields, we can address both the $\ket{0}\rightarrow\ket{-1}$ and $\ket{0}\rightarrow\ket{+1}$ NV ground state spin transitions, producing two resonances at frequencies $f_-=D-\gamma_{\rm NV}B_{\rm NV}$ and $f_+=D+\gamma_{\rm NV}B_{\rm NV}$ as shown in Fig.~\ref{SI_odmr}, where $D$ is the zero-field splitting, $\gamma_{\rm NV} = 28.033(3)$~GHz/T is the NV gyromagnetic ratio, and $B_{\rm NV}$ is the total magnetic field projection along the NV axis. These resonances were fit by Lorentzian functions with bounded frequencies, amplitudes, and widths. Taking the difference of the two frequency maps gives the desired $B_{\rm NV}$ map. In practice we subtract the bias field and any background variation (e.g. due to variations in laser intensity or microwave driving artificially altering the fit frequencies) across the field of view to obtain the signal only due to CCPS flakes. As shown in Fig.~\ref{SI_odmr}, a map of $D=\left(f_+ + f_-\right)/2$ can also be obtained, which is seen to vary across the field of view due to crystal strain, which is sometimes intrinsic to the crystal but more commonly in our samples is due to the Al grid deposition. These features are typically large in magnitude compared to the magnetic signal from target flakes, so removing them by addressing both transition frequencies is important.

For measurements that required the application of a stronger bias field ($|B|\geq 60$~mT) such as those in Fig. 3 of the main text, the frequency of the $\ket{0}\rightarrow\ket{+1}$ transition was out of the range of the amplifier used and was thus inaccessible. This meant that magnetic images had to be obtained using only one ODMR peak. Strain features could still be removed by subtracting a $D$ map obtained at low field over the same region. The $\ket{0}\rightarrow\ket{-1}$ transition also becomes inaccessible past $\approx 75$~mT as the frequency becomes too low for the amplifier, but is able to be addressed again at around 1400~G as past $\approx 100$~mT the NV ground state undergoes an anticrossing and the relevant transition frequency begins to increase with applied field. The $\ket{0}\rightarrow\ket{-1}$ transition finally passes the high-end limit of our amplifier at around 300~mT. These limitations meant our variable-field measurements were fairly coarse, however in principle the technique can be extended to an arbitrary range of fields provided the technical requirements are met.

\section{Data processing and analysis}
Compared to the case of out-of-plane magnetization, where magnetization reconstruction from measured stray field is effective~\cite{Broadway2020,Broadway2020a}, systematically reconstructing in-plane magnetization is less trivial due to the uncertainty in magnetization direction and nontrivial ways the border features from different domains may intersect or overlap. Because of this, particularly for the more complex regions, we prefer to compare with simulations to quantify $M_{\rm s}$ as outlined in Sec. \ref{si: sims} (and as also preferred in Ref.~\cite{Fabre2021}). However, we can also reconstruct individual magnetic field components to gain qualitative information about the magnetization direction, as was argued in the main text. Here the large field of view of our measurements is advantageous in avoiding truncation artefacts~\cite{Scholten2021}. 
\subsection{Magnetic field reconstruction}
\begin{figure}
\includegraphics[width=0.8\textwidth]{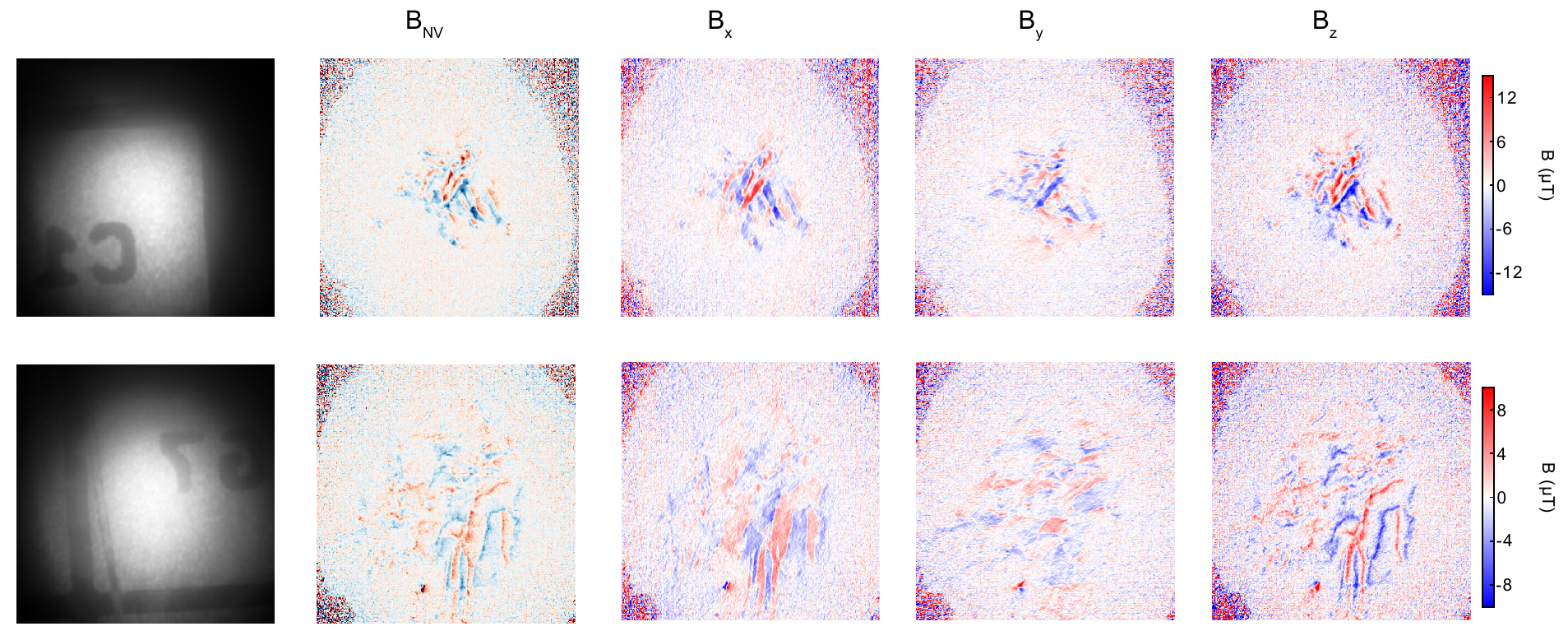}
\caption{From left to right: PL images, $B_{\rm NV}$ images, and reconstructed $B_x$, $B_y$, and $B_z$ images for the low-field images Fig. 2(c) (top) and Fig. 1(d) (bottom). In both cases the reconstructed fields are consistent with magnetization confined to the $ab$ plane.}
\label{sifig:inplane}
\end{figure}

In this section we present the reconstructed magnetic fields from selected $B_{\rm NV}$ images shown in the main text. First, we show in Fig.~\ref{sifig:inplane} that low-field magnetic images shown in Fig. 1 and 2 of the main text are consistent with purely in-plane magnetization as claimed. Note specifically that the in-plane stray field detected within a flake's boundaries is oriented oppositely to the direction of its in-plane magnetization, and so the reconstructed $B_x$ map for the Fig. 1 flake in particular matches well with the domain structure proposed in Fig. 1(e). The reconstructed field components are less revealing for the Fig. 2 region due to the smaller domains present, however it can still be seen that the $B_z$ map contributes the sharpest border-like features while the in-plane components give the fields internal to domains. In both cases the measurement field of view extends far beyond the regions of interest and so we can be confident that the reconstructed field maps are relatively free from truncation artefacts. 

\begin{figure}
\includegraphics[width=0.5\textwidth]{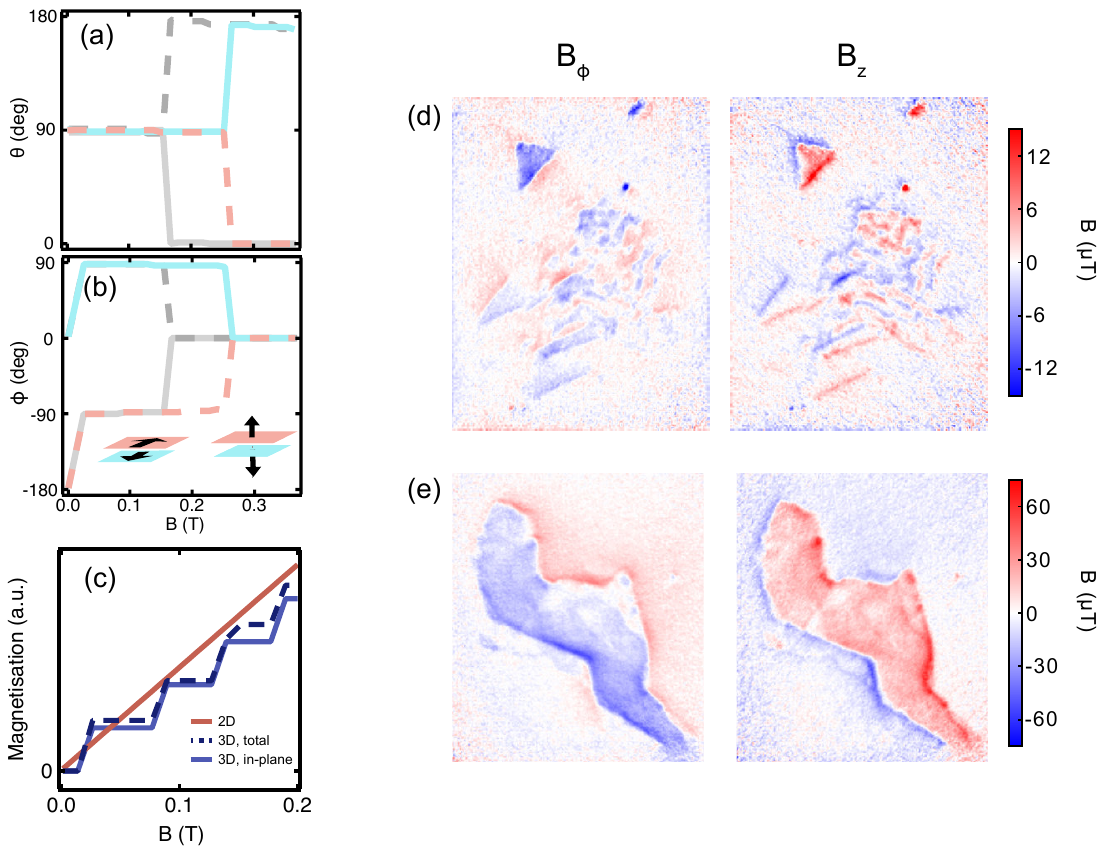}
\caption{(a), (b) Respectively, the polar and azimuthal angles corresponding to the lowest energy configuration of sublattice spins for a two-layer system, computed using the three-dimensional micromagnetic model described in text. Grey curves show the configuration when a small surface anisotropy $K_z = 100$~J/m$^3$ is introduced. (c) Comparison between the magnetization predicted by the three-dimensional (no surface anisotropy) and two-dimensional models, showing close agreement. Note that the small out-of-plane component over this range of fields is due to a slight canting, not the spin-flop that occurs past 0.25~T. (d) Reconstructed fields for the $B_{\rm ap} = +0.2$~T image from Fig. 3(a) of the main text, showing the magnetization remains mostly confined to the $ab$ plane at this field, especially for the thinner regions. (e) Reconstructed fields for the $B_{\rm ap} = +0.2$~T image from Fig. 3(b) of the main text, showing significant canting out of the $ab$ plane.}
\label{sifig:highfield}
\end{figure}

We next consider the field-dependent series of data from Fig. 3 of the main text. For both regions, we see in Fig.~\ref{sifig:highfield}(d),(e) that at high fields the field reconstruction suggests internal $z$ fields are measured in thick flakes, consistent with a slight canting of magnetization towards the applied field axis. In light of this, one may question whether the purely two-dimensional model used to analyse the data in the main text is limiting. At the cost of greater computational stress, we can extend this model to three dimensions and again compute the lowest energy configuration of sublattice spins under the assumptions of the model. We assume that the antiferromagnetic exchange acts in the same way regardless of moment orientation and include a thin film shape anisotropy term $K_{\rm sh} = \sum_{i=1}^{2}\frac{1}{2} \mu_0 M_{\rm s} \cos{\theta_i}$ to describe the easy-plane behaviour, which agrees with the values quoted in Ref.~\cite{Qi2018}. In Fig.~\ref{sifig:highfield}(a) and (b) we plot the lowest energy configurations of the moments $\bm{m}_1$ and $\bm{m}_2$, taking a two-layer model for computational simplicity, with the applied field lying along the NV axis ($\theta = 55 \degree$, $\phi = 0 \degree$ as in the main text). In this simple model, we see that a secondary spin-flop transition is predicted when the $z$ projection of the applied field exceeds the approximate threshold $\sqrt{J_{\rm AFM}K_{\rm sh}}$, leading to a net out-of-plane magnetization. Figure~\ref{sifig:highfield}(c) shows, however, that there is close agreement between the magnetizations predicted by the two-dimensional and three-dimensional models (two layers considered in both cases). Since the two-dimensional model is much less computationally intensive and thus allows both smoother predictive curves to be obtained and odd-layered stacks to be modelled, we prefer it for the analysis presented in the main text.

This threshold for the second spin-flop transition reduces if a weaker anisotropy favours out-of-plane magnetization (as shown by the grey curves in Fig.~\ref{sifig:highfield}(a),(b) which takes a value $K_z = 100$~J/m$^3 \approx 0.01 K_{\rm sh}$), and even if it is not strong enough to render the out-of-plane configuration energetically favourable at low field it may be sufficient to stablise a metastable state~\cite{Bogdanov2006}. The most likely source of such an anisotropy in the present system would be a surface anisotropy initially affecting primarily the outermost layer(s), then propagating through the bulk material with increasing field. This could be an intrinsic effect that the CCPS monolayer is always susceptible to, or could be induced (or exacerbated) by degradation of the outermost layers due to air exposure. Given the range of variable behaviour observed between different CCPS flakes in our study it is reasonable to conclude that the effective $K_z$ introduced by sample preparation or degradation can be highly variable. 

As with the two-dimensional model, the two-layer model will be inaccurate for sufficiently thin odd-layered flakes whose behaviour will be impacted by the Zeeman term arising from the uncompensated layer. Given that the two-dimensional model predicted the in-plane spin-flop to take place at fields beyond our measurement range for flakes 5, 7, etc. layers thick, it is unsurprising that magnetic field reconstruction shows that these flakes retain purely in-plane magnetization over the full range of fields probed [see Fig.~\ref{sifig:highfield}(d)]. This also shows that it is possible to mechanically exfoliate few-layer CCPS flakes with minimal surface anisotropy meaning that this material remains a promising candidate for realising robust, easily-switchable in-plane magnetization.
\subsection{Simulations}
\label{si: sims}
The simulations used to compare with experimental data in Fig. 1 and 2 were obtained by defining uniformly magnetized domains of shapes matching the experimental patterns. In both cases, the magnetization was forced along the direction of a previous training field and qualitative agreement was found with experiment. By including an NV-flake standoff of 300~nm in the simulation and accounting for the optical diffraction limit through an additional convolution with a Gaussian function, the only remaining free parameter is the magnitude of the magnetization within the domains. Therefore, we can estimate the spontaneous magnetization $M_{\rm s}$ of our CCPS flakes by comparing simulation with experiment.

\begin{figure}
\centering
\includegraphics[scale=1]{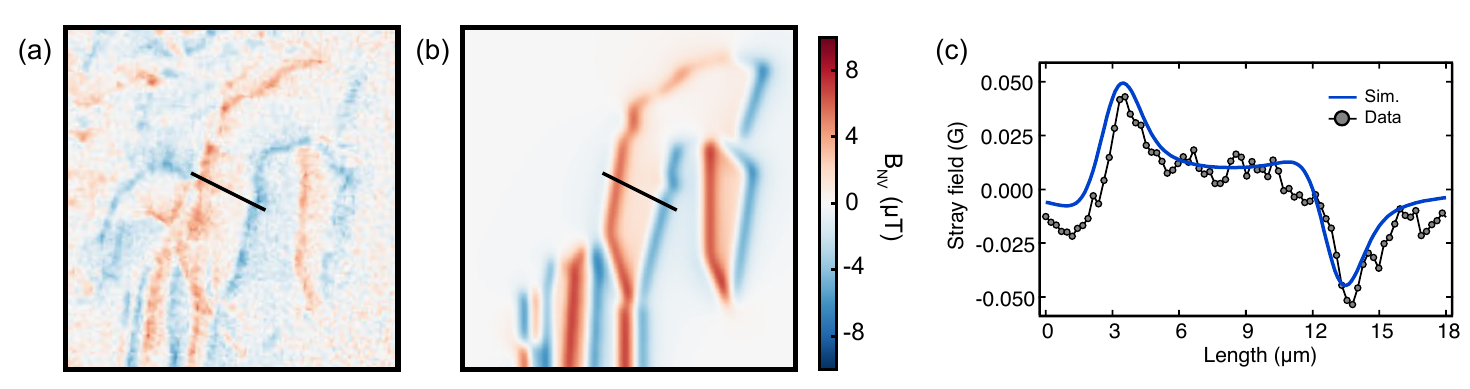}
\caption{(a) Experimental $B_{\rm NV}$ map from Fig. 1(d). (b) Simulated $B_{\rm NV}$ map from Fig. 1(f). (c) Comparison of linecuts taken across the lines marked on (a) and (b).}
\label{sifig: sims}
\end{figure}

The width of the Gaussian function to be convolved with the simulated field was set to match the width of stray field features in a given image. The optical diffraction limit can be as low as $\approx 500$~nm in our setup, where it is primarily limited by optical aberrations occurring at the various interfaces, however may also vary for different regions across the diamond due to variable focus or sample tilt. The standard deviation of the Gaussian function was 500~nm for the Fig. 2 simulation and 750~nm for the Fig. 1 simulation. In both cases, the value of $M_{\rm s}$ that then gave matching stray field amplitudes was $\approx 3.5~\mu_B$/nm$^2$. The agreement is shown in Fig.~\ref{sifig: sims}(c) where linecuts taken across the experimental image (grey dots) and the simulated image (blue line) match well. 

\begin{figure}
\includegraphics[width=0.5\textwidth]{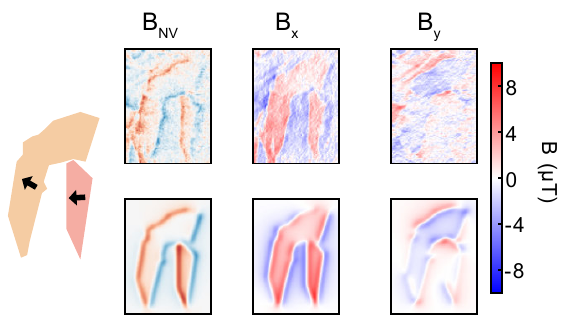}
\caption{Comparison between experimental $B_{\rm NV}$ and reconstructed $B_x$ and $B_y$ maps (top) and the simulated fields along these axes (bottom) when taking the direction of magnetization in the two large domains to be as illustrated in the schematic (left).}
\label{sifig:simangle}
\end{figure}

By considering the reconstructed fields in Fig.~\ref{sifig:inplane} (reproduced in the top row of Fig.~\ref{sifig:simangle}) we can also arrive at a more accurate simulation for the two largest domains. As the CCPS crystal has six-fold symmetry and previous reports have suggested a preferred direction of magnetization along the $a$ axis~\cite{Maisonneuve1995}, we can consider whether a $30\degree$ rotation of the magnetization direction away from the direction of the training field may better represent the experiment. In Fig.~\ref{sifig:simangle} we can see that tilting the magnetization direction in the left domain does produce a more faithful simulation. The right domain is well represented by magnetization purely in the $-x$ direction. In principle, this procedure can help identify the direction of in-plane magnetization and aid quantitative magnetization reconstruction~\cite{Broadway2020a}. 

\begin{figure}
\includegraphics[width=0.5\textwidth]{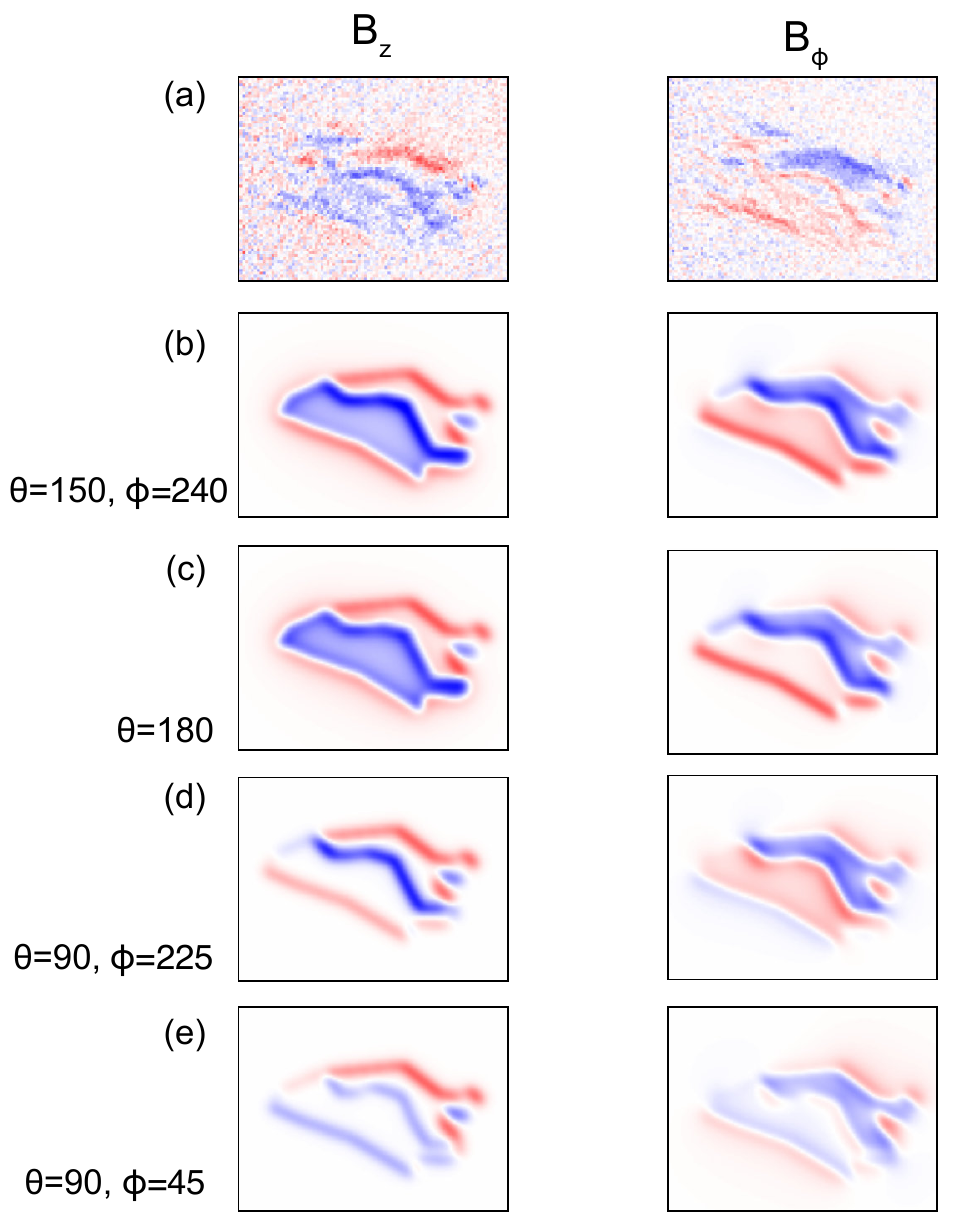}
\caption{(a) Reconstructed $B_z$ (left) and $B_{\phi}$ (right) components for the $B_{\rm ap} = -15$~mT measurement, Fig. 4(c) of the main text. (b) Simulated $B_z$ and $B_{\phi}$ fields taking magnetization along $\theta = 90\degree$, $\phi = 45\degree$ for the top right domain and $\theta = 150\degree$, $\phi = 240\degree$ for the bottom right. (c) As (b) but with the bottom right domain taking $\theta = 180\degree$. (d) Bottom domain $\theta = 90\degree$, $\phi = 225\degree$, magnetization half the magnitude of the top left domain. (e) $\theta = 90\degree$, $\phi = 45\degree$, magnetization half the magnitude of the top left domain.}
\label{sifig:outplane}
\end{figure}

We can also justify the interpretation of Fig. 4 of the main text using simulation. Figure~\ref{sifig:outplane} compares the reconstructed $B_z$ and $B_{\phi}$ components from Fig. 4(c) with some candidate magnetization directions for the bottom left domain (keeping the top right domain constant; magnetized in-plane along the angle $\phi = 45\degree$). In this case, due to the large uncertainty in the direction of magnetization and precise domain structure, we do not seek quantitative agreement and so the magnitude of the stray fields and input magnetizations are somewhat arbitrary (we used $M_{\rm s} = 2.7~\mu_B$/nm$^2$ and convolved with a Gaussian with standard deviation 750~nm). Instead we can see that to reproduce the qualitative feature of significant $B_z$ fields on the interior of the bottom left domain, it is necessary for the magnetization to have a large $z$ component, as claimed in the main text. The relatively significant internal $B_{\phi}$ field also suggests an in-plane component [e.g. magnetization along $\theta = 150\degree$, $\phi = 240\degree$ as in Fig.~\ref{sifig:outplane}(b)], although a pure out-of-plane magnetization cannot be ruled out [Fig.~\ref{sifig:outplane}(c)]. It is not possible to produce $B_z$ and $B_{\phi}$ fields that match those reconstructed from experiment using purely in-plane magnetization [Fig.~\ref{sifig:outplane}(d),(e)], even when tweaking the magnetization of the bottom left domain relative to the top right (we reduced the magnetization by a factor of two here). The relative magnitudes of the stray field features are consistent with experiment using the same magnetization for both domains in Fig.~\ref{sifig:outplane}(b),(c), supporting the interpretation that the signal is due to an uncompensated, ferromagnetic monolayer in both cases. The simulation does produce larger stray field along the domain border than is seen in experiment, however this is explained by the simplicity of the two-domain model; in reality the structure is likely more complex than this, including the possibility of micro-domains below our measurement resolution along this border.

\subsection{Magnetization measurements}
Having established that simulated uniformly magnetized domains reproduce our experimental results well, and that linecuts can be used to compare the two and infer the magnetization of a flake, we move to describing how extended data sets were used to build up Fig. 3(c) of the main text. The magnetization for each image was inferred from a linecut taken across the flattest part of the flake (and hence the region whose height measured by AFM will most accurately reflect the total number of layers). Due to the various sources of error mentioned in the main text, our measurements cannot precisely identify the monolayer $M_{\rm s}$, however relative changes are measurable. The maximum value of a linecut is linear in the magnetization, and so comparing these maxima gives a relative measure of magnetization. Where possible, to control for varying impact of delocalized PL across the diamond (e.g. due to optical aberrations altering laser delivery or PL collection, or due to the inhomogeneous Al grid introducing variable reflections) a maximum signal exhibited by a thin CCPS flake at low field was used as a baseline ($M=M_{\rm mono}$). In general, the following additional considerations need to be taken into account in order to correctly infer the magnetization scaling:
\begin{itemize}
\item Reduced measured stray field due to misalignment between the direction of magnetization and the direction normal to the flake edge.
\item Apparently reduced magnetization due to significant thickness of the flake.
\end{itemize}
The latter factor was particularly important to consider for the 300 nm thick flake (Fig. 3(b) main text), whose thickness was comparable to the measurement standoff. This was taken into account by simply considering the reduction in stray field measured over the NV measurement plane (at a depth 300~nm) when the magnetization is spread evenly throughout a 300~nm slab versus concentrated on the diamond surface. This is a factor of approximately 1.4, by which we multiplied the relative magnetizations measured. 
\subsubsection{Error propagation for the measurement of $J_{\rm AFM}$}
In the analysis surrounding Fig. 3 of the main text, a measurement of the interlayer antiferromagnetic exchange coupling was made using the data obtained from the 300~nm flake (black points, orange curve). The theoretical scaling of magnetization with applied field as predicted by the micromagnetic model is approximately linear over the range of measurement fields used, and the experimental data is well fit by a linear function with an error of $\approx 2\%$ in the gradient. More significant uncertainties come from the determination of the layer number $N$ and the monolayer saturated magnetization $M_{\rm s}$. Combining the uncertainty in the AFM measurement and the thickness of a single layer (noting a small discrepancy between the monolayer step height noted in SI-II C and literature values~\cite{Liu2016,Lai2019}), we came to a relative error of 0.15 for $N$. In the micromagnetic model, we took the theoretical value for $M_{\rm s}$ [with the spin-only value for Cr(III)]. However, there is reason to believe that the actual $M_{\rm s}$ may be smaller than the theoretical value, especially considering the sample temperature under measurement conditions may exceed 5~K. For this reason, we took our experimental value as a lower bound and the spread between it and the theoretical value as the error in $M_{\rm s}$.

To obtain an uncertainty in $J_{\rm AFM}$ we then followed a procedure similar to  that used by Fabre et al~\cite{Fabre2021}. For each parameter $p$ with an uncertainty $\sigma_p$ we found the relative uncertainty introduced in the determination of $J_{\rm AFM}$ $\epsilon_p$ by calculating
\begin{equation}
\epsilon_p = \frac{J_{\rm AFM} (p+\sigma_p) - J_{\rm AFM} (p-\sigma_p)}{2J_{\rm AFM} (p)},
\end{equation}
where $J_{\rm AFM}(p \pm \sigma_p)$ is the value that gives a fit consistent with the data when $p$ is varied to its extremal value, with all other parameters held constant. The total uncertainty was evaluated as the weighted average of these uncertainties and we arrive at the value quoted in the main text, $J_{\rm AFM} = (1.19 \pm 0.56) \times 10^{-4}$~J/m$^2$. 
\section{Additional data}
\subsection{Coercivity measurements}
\begin{figure}
\includegraphics[width=0.9\textwidth]{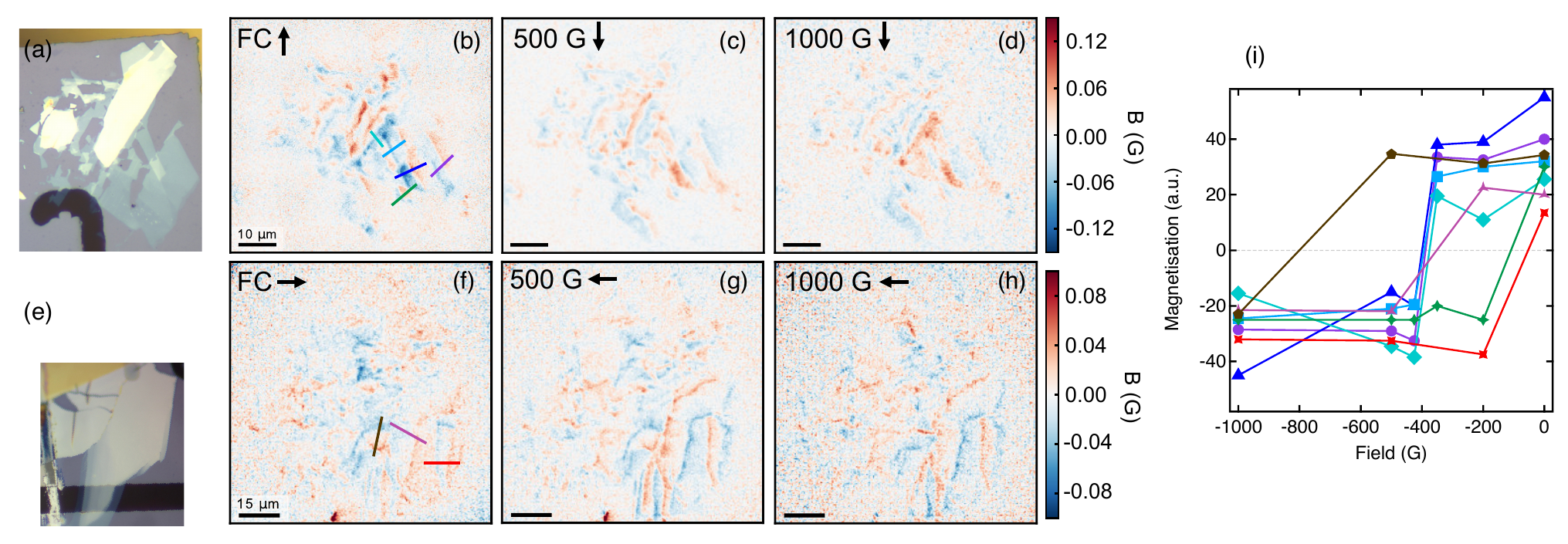}
\caption{(a) Optical image of Fig. 2 region. (b) Magnetic image of C1 region taken at 60~G following field cooling under 1~T in the ``$-y$" direction. (c), (d) As (b) but following the application of 500 and 1000~G respectively in the ``$+y$" direction at 4~K. (e)-(h) As (a)-(d) but for the Fig. 1 region and with fields in along the ``$x$" direction. (i) Summary of full magnetization data, inferred from the linecuts marked on panels (b) and (e). The sign of the magnetization is arbitrary but is seen to flip at variable fields but most commonly near 400~G.}
\label{sifig: coercivity}
\end{figure}

In order to characterize the coercivity of odd-layered ferromagnetic CCPS flakes, flakes were field cooled from above $T_N$ ($T=60$~K) under fields of 1~T along a chosen in-plane axis. Magnetic images of the field cooled state were obtained at 5~K under a bias field of 6~mT [which was not large enough to switch magnetization in thin flakes, see Fig.~\ref{fig: MH}(a)]. A field of a given magnitude was then applied in the opposite direction to the field cooling at 5~K, before the resulting magnetization was again imaged. This was repeated with increasing switching fields until nearly all features were observed to be switched (up to 0.1~T in practice). The desired field was typically reached within a minute and then the maximum field was maintained for $\approx 10$~s before being reduced back to zero. Reversal of a domain is indicated in our images by a reversal in sign of opposite domain walls, although the structure is usually complex. Magnetization reversal was quantified by taking the maximum values from the linecuts shown on Fig.~\ref{sifig: coercivity}(b),(f) and designating positivity or negativity based on the sign of the features on the field cooled image. These results are summarised in Fig.~\ref{sifig: coercivity}(i). In principle these linecuts could be compared to simulation individually to obtain a direct measure of domain magnetization, however given that the stray field magnitude will be affected by factors such as the domain size (often small here) and the precise direction of magnetization, we just plot the maximum stray field directly in Fig.~\ref{sifig: coercivity}(i) as a proxy for magnetization.

It can be seen that, for most flakes across both regions examined, the coercive field is typically $\approx 40$~mT, although some regions switch at higher or lower fields than this. It is possible that strain-induced anisotropy is dominant in smaller flakes, so these values could be flake-specific rather than intrinsic. As similar switching fields are measured for flakes with a range of thicknesses (between 5 and over 40 layers), we do not see any evidence for the coercivity varying with the number of layers. This is likely because the inter-layer $J_{\rm AFM}$ is sufficiently weak compared to the intra-layer coupling. 

A spread in the magnitudes of fields in Fig.~\ref{sifig: coercivity}(i) is (in addition to the factors mentioned above) likely reflective of the variation of domain boundaries relative to the direction of magnetization. Straightforwardly, for a given magnetization and a rectangular domain, the stray field at an edge varies as the cosine of the angle the magnetization makes with its normal. Simulation confirms this remains true when including a standoff and convolving with the measurement spatial resolution. Therefore, considering the six-fold symmetry of the CCPS lattice, we might assume that the spread in inferred magnetization may be roughly $\cos{\left(30\degree\right)} M_{\rm s}$, or a spread of about 15\%, which is consistent with our data. As shown in Fig.~\ref{sifig:simangle}, the direction of magnetization is not necessarily locked to the direction of the training field. 

\subsection{Magnetic signal in trilayer regions}
Here we present evidence for magnetic order being maintained down to trilayer thickness in CCPS. Two trilayer regions were identified, one each in the areas the subject of Fig. 1 and Fig. 2 of the main text. 

\begin{figure}
\includegraphics[width=0.9\textwidth]{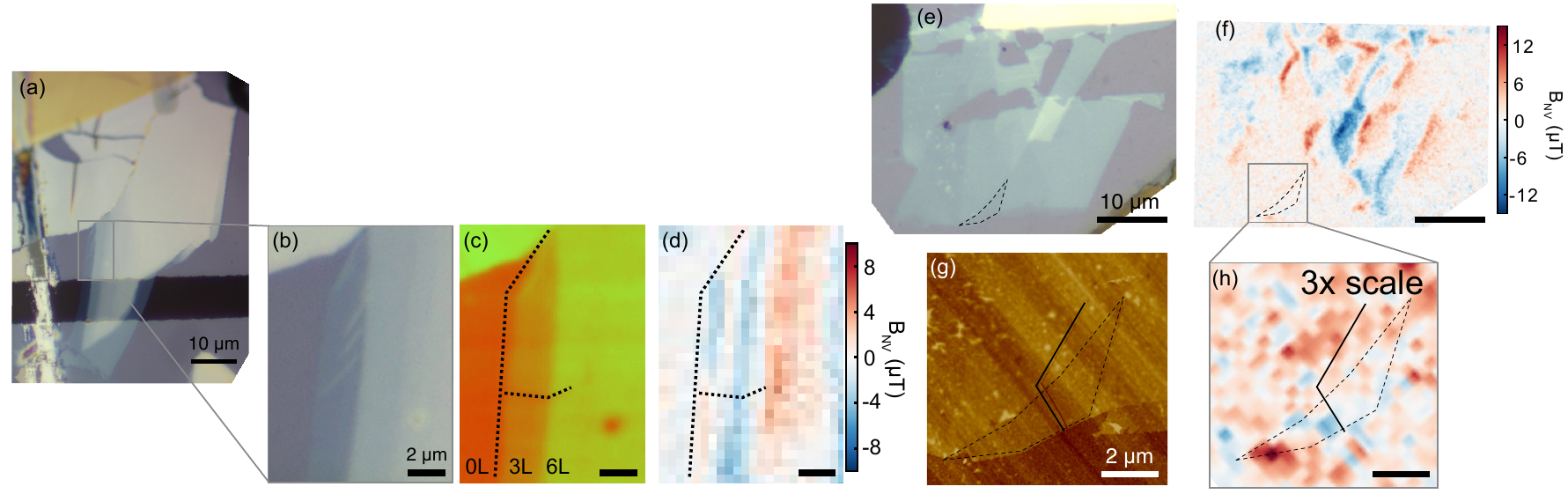}
\caption{Observation of ferromagnetic signal at low field in two trilayer CCPS flakes. (a) Optical image of the region from Fig. 1. (b) Zoom-in on the grey box in (a). (c) PSI image of the region from (b). (d) $B_{\rm NV}$ image of the same region. Dotted lines show the edge of the 3L region and a fracture visible in the PSI image, which is seen to correlate with a boundary between oppositely-oriented domains [see also, Fig. S5 and the simulation Fig. 1(f)]. (e) Optical image of the region from Fig. 2, with a trilayer region outlined (as identified in Fig. S4). (f) A $B_{\rm NV}$ image of the region from (e). (g) AFM image of the trilayer flake in (e). The solid black line highlights a large crack. (h) Zoomed $B_{\rm NV}$ image of the grey box region in (f), shown on a smaller field scale so that a weak magnetic domain in the bottom section of the trilayer is evident. The domain's extent is seen to correlate with the location of the crack visible in (g).}
\label{SIfig:trilayermag}
\end{figure}

The clearest signal was found in the region from Fig.~\ref{sifig:trilayer}, optical and PSI images reproduced in Fig.~\ref{SIfig:trilayermag}(a)-(c). A trilayer strip about as wide as our measurement resolution ($\approx 1$~$\upmu$m) is resolvable in our $B_{NV}$ image [Fig.~\ref{SIfig:trilayermag}(d)]. The PSI image for this region [Fig.~\ref{SIfig:trilayermag}(c)], which reveals a fracture across the flake that is not visible optically. We see in the $B_{NV}$ image that this coincides with a border between oppositely oriented domains (stripes of opposite coloration, see also the simulation Fig. 1(f) of the main text). Considering the  order of fields applied before taking this image ($-1$~T along the $x$ axis, measurement under $B_{\rm ap} = 6$~mT along $\hat{u}_{NV}$, $+50$~mT along the $x$ axis, measurement under $B_{\rm ap} = 6$~mT along $\hat{u}_{NV}$), we can understand this fracture as a domain wall pinning site. The upper section of the trilayer may, for example, be more highly strained, leading to a greater resistance to magnetization reversal. 

The magnetic signal measured from the trilayer is slightly lower in magnitude than some other features in this image, however from the simulation Fig. 1(f) we can understand this solely as a size effect; the domain size is of the order of the measurement resolution and so the signal arising from opposite edges leads to some amount of cancellation. Therefore our results are consistent with a uniform magnetization in the trilayer of the same magnitude as in magnetic domains observed in thicker flakes. 

Another trilayer exhibiting magnetic signal is found in the region from Fig. 2 of the main text (thickness analysis in Fig.~\ref{sifig:psiC1}), highlighted in Fig.~\ref{SIfig:trilayermag}(e); the corresponding $B_{\rm NV}$ image is shown in Fig.~\ref{SIfig:trilayermag}(f). An AFM image [Fig.~\ref{SIfig:trilayermag}(g)] shows a significant crack through the trilayer which possibly explains why clear magnetic signal is not visible in Fig.~\ref{SIfig:trilayermag}(f). A zoomed-in $B_{\rm NV}$ image [Fig.~\ref{SIfig:trilayermag}(h)] agrees with this picture: low-level stray field signal consistent with an in-plane magnetic domain is present in the half of the trilayer below the crack while none is visible above it. The size of these individual regions is marginally above our spatial resolution, so long-range magnetic order should in principle be measurable and the small signal is not explained by size effects alone as in the previous case. Instead, it is likely that the exfoliation process has incurred a significant enough level of damage to disrupt the formation of ferromagnetic domains in this trilayer (possibly leading to domains below the measurement resolution or with a substantially lowered $T_c$ compared to the bulk material). Further, this small magnetic signal was only visible occasionally in our measurements: the formation of a large enough domain to measure may have been dependent on the training field(s) applied and/or a higher than usual level of acquisition was required for this signal to be above noise. Nevertheless, the appearance of this signal does imply that a pristine CCPS trilayer can easily sustain long-range magnetic order.

Magnetic signal was not observed in either of the monolayer regions highlighted in Fig. 2(b). Again, at least one of these regions is comfortably larger than our measurement resolution although given our observation that long-range magnetic order can be disrupted by deposition- or exfoliation-induced damage in 3L and 5L flakes, it is possible that the monolayers were simply not well enough preserved. 
\subsection{Measurements versus temperature}
\begin{figure}
\includegraphics[width=0.8\textwidth]{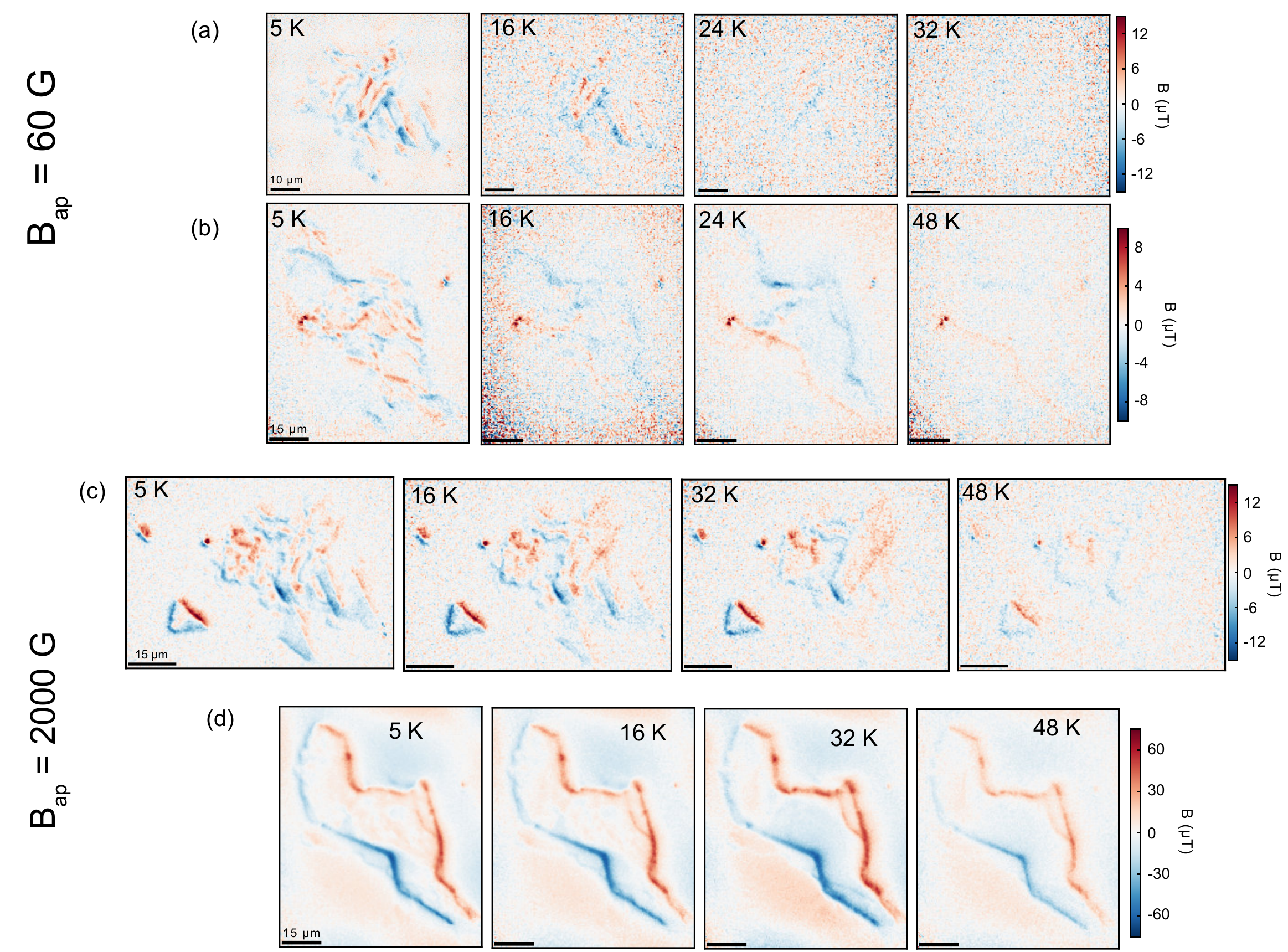}
\caption{Series of measurements taken over a range of temperatures: (a) thin region (Fig. 2 and Fig. 3(a) main text), low field; (b) thick flake (Fig. 3(b) main text), low field; (c) thin region, high field; (d) thick flake, high field.}
\label{SIfig:temp}
\end{figure}
The temperature dependence of the magnetization of CCPS flakes at both low and high applied fields (60~G and 2000~G respectively) was measured by raising the sample temperature above the cryostat base temperature (4~K) using a heater located directly below the sample stage~\cite{Lillie2020}. The temperature was read out using a sensor located near the heater; as the thermal contact between the sensor and heater is better than that between the heater and the sample, the measured temperature systematically overestimates the actual sample temperature. Previous calibration under similar experimental conditions showed that this offset can be corrected for by multiplying the measured temperature by a constant factor 0.8 when the heater is active~\cite{Broadway2020} and this correction has been applied in all quoted temperatures below. Additionally, although the measured temperature without the heater being active was always $<4.5$~K, it is possible that the application of laser and microwave pulses during an ODMR protocol could raise the sample temperature above its nominal value~\cite{Lillie2020}. By including the Al laser shield in our diamond sensor design we are able to limit the impact of laser heating, and heating due to microwave delivery was determined to be negligible with our pulsed ODMR protocol \cite{Lillie2020}. We therefore estimate that the true sample base temperature in our measurements is 5~K.

Beginning with the low-field measurements of thin flakes [Fig.~\ref{SIfig:temp}(a)], we see that the ferromagnetic signal reduces in magnitude until it disappears completely past the N\'eel temperature $T_N=31$~K, consistent with previous studies~\cite{Maisonneuve1995,Kleemann2011}. A natural question to ask is if the critical temperature varies with film thickness as for other van der Waals magnets, however the relatively coarse temperature control available in our experiment was not sufficient to investigate this properly, and so we leave this as an open question for future work. 

Fig.~\ref{SIfig:temp}(b) shows the low-field temperature dependence of the thickest flake studied, showing that the complex stray field patterns observed at 5~K yield to well-defined edge features at 24~K. We believe that this is evidence of temperature dependence of the spin-flop transition, specifically that it takes place at lower fields as temperature increases. This flake is thick enough that the small paramagnetic magnetization is measurable past $T_N$, up to a temperature of 48~K.

Moving to the high-field measurements, Fig.~\ref{SIfig:temp}(c) shows that, as in the low-field case, the signal from thin flakes rapidly decreases near $T_N$ as expected of a ferromagnet. The fields from the thick regions, while of a similar magnitude to the ferromagnetic signal at 5~K, do not decrease in magnitude as rapidly with increasing temperature. This behaviour is expected of an antiferromagnet in the spin-flop phase, supporting the analysis surrounding Fig. 3 of the main text. Due to the stronger applied field, the paramagnetic signal past $T_N$ is measurable in the thicker flakes now.

Finally, Fig.~\ref{SIfig:temp}(d) shows the high-field behaviour of the thick flake. As with the thicker regions in Fig.~\ref{SIfig:temp}(c), this flake is in a spin-flop configuration at this field and so the magnitude of the stray field signal does not change much with temperature, however towards $T_N$ we do observe a slight increase in magnetization consistent with the weakening impact of the antiferromagnetic interlayer exchange coupling. Again this is consistent with expectation and previous measurements of bulk and powdered CCPS samples which show a small cusp in magnetic susceptibility at $T_N$~\cite{Colombet1982,Kleemann2011}. This increase is further evidence that the behaviour at lower temperatures is described by a spin-flop transition and not simply a paramagnetic signal.

\subsection{Anisotropy}
\begin{figure}
\centering
\includegraphics[width=0.9\textwidth]{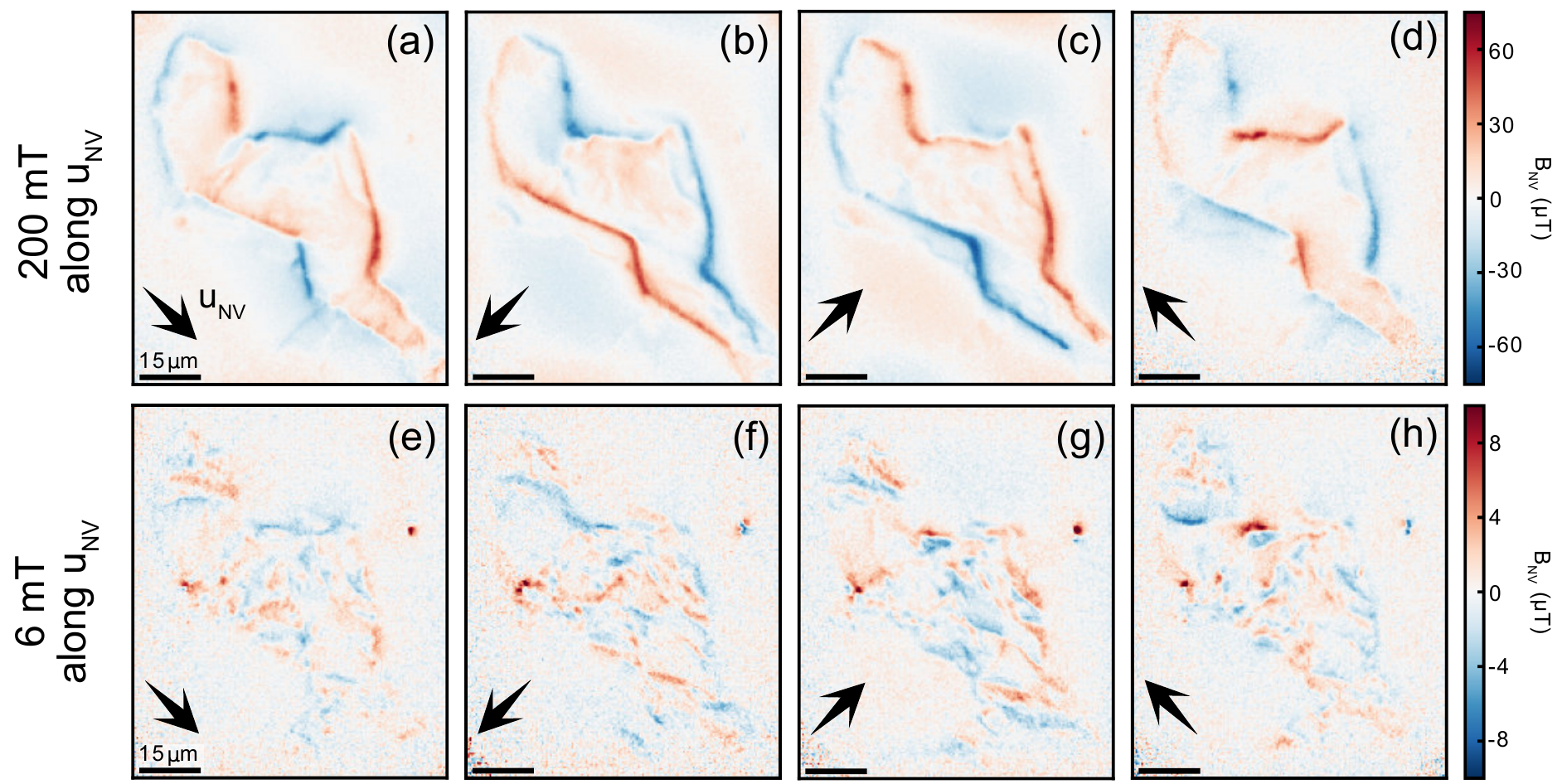}
\caption{Lack of in-plane anisotropy. (a), (b), (c), and (d) show the signal obtained when measuring a thick flake at a field of 200~mT along the NV families aligned at 135, 45, 225, and 315 degrees to the x-axis respectively, where $u_{\rm NV}$ is defined to point out of the page in all cases. (e), (f), (g), and (h) show the corresponding images taken a negligible bias field of 6~mT. We see some evidence of memory of the high-field spin configuration but overall a more fragmented domain structure.}
\label{fig: anisotropy}
\end{figure}

As the model used to obtain Fig. 3(c),(d) of the main text assumes negligible magnetic anisotropy, it is sensitive to the addition of an aniostropy term. Qualitatively, non-zero anisotropy will increase the spin-flop threshold ($H_{\rm SF} \approx \sqrt{J_{\rm AFM}K}$) and so the observation of a near-zero-field spin-flop in thick flakes places an upper bound on the magnitude of any intrinsic in-plane magnetocrystalline anisotropy. In Fig.~\ref{fig: anisotropy} we present further evidence for there being negligible in-plane anisotropy in the large, thick flake.  
Applying $B_{\rm ap} = 200$~mT along the four NV axes (aligned at 45, 135, 225, and 315 degrees from the $x$-axis, and choosing the $z$-component of the applied field to be out of the page in all cases), we see that the flake readily magnetises along that axis, providing distinct stray field patterns in each case. Upon decreasing the applied field to 6~mT (along the same axis), we see in each case that the direction of the overall magnetization of the flake (given by the sign of the stray field along the flake borders) is set by the alignment at 200~mT rather than any in-plane anisotropy. The internal structure re-emerges at the lower field and is distinct in each case with some similarities, indicating that while the shape of the defects present in the stack is important, the intrinsic magnetocrystalline anisotropy is not enough to overcome the Zeeman term. As AFM images are only sensitive to the topography of the top layer, where a level of buckling is visible in some areas, we are unable to correlate the internal coloration (at both low and high fields) to specific flake defects, however it is expected that various stacking faults and fractures are responsible. At low field, they act as pinning sites to separate domains while at high field they still set the boundaries for internal variation in signal (much weaker than the dominant edge features). 

The near-free rotation of magnetism in-plane at fields much lower than those required to appreciably rotate moments out of plane~\cite{Qi2018} implies an XY-Heisenberg model describes CCPS at low fields, as is also the case in CrCl$_3$~\cite{Mcguire2017}. Note, however, that this behaviour is only dominant so long as a flake is thick enough that the Zeeman term from any uncompensated odd layer is relatively insignificant in working against the tendency to spin-flop and is not expected to be as applicable for thin odd-layered stacks.

This result indicates that intrinsic magnetocrystalline in-plane magnetic anisotropy in CCPS is negligible, supporting the above analysis. Some level of anisotropy is required for magnetic order to be sustained over large length scales, and has previously been observed in CCPS using neutron scattering~\cite{Maisonneuve1995}, however its magnitude has not been measured. As for CrCl$_3$, it appears that this anisotropy is large enough to sustain magnetic order, but small enough to allow effectively free rotation of the magnetic moments with weak magnetic fields. 

\end{document}